\documentclass[twocolumn,tighten]{aastex631}
\usepackage{color}
\usepackage{comment}
\newcommand{\scm}{cm$^{-2}$}	% per cm-squared
\newcommand{\ccm}{cm$^{-3}$}	% per cm-cubed

\newcommand{\kms}{km~s$^{-1}$}
\newcommand{\HII}{H{\sc ii}\,}
\newcommand{\HI}{H{\sc i}\,}
\newcommand{\co}{$^{12}$CO}

\accepted{January 24, 2026}

\shorttitle{Gas Kinematics and Cosmic-Ray Acceleration in the Gamma-ray SNRs W41 and G22.7–0.2}
\shortauthors{Murase, Sano and Matsubara et al. (2026) }

\begin{document}

\title{Gas Kinematics and Cosmic-Ray Acceleration in the Gamma-ray SNRs W41 and G22.7–0.2}

\author[0000-0002-9552-3570]{T. Murase}
\affiliation{{Faculty of Engineering, Gifu University, 1-1 Yanagido, Gifu 501-1193, Japan}}

\author[0000-0003-2062-5692]{H. Sano}
\affiliation{{Faculty of Engineering, Gifu University, 1-1 Yanagido, Gifu 501-1193, Japan}}
\affiliation{{Department of Intelligence Science and Engineering, Graduate School of Natural Science and Technology, Gifu University, 1-1 Yanagido, Gifu, 501-1193 Japan}}
\affiliation{{Center for Space Research and Utilization Promotion (c-SRUP), Gifu University, 1-1 Yanagido, Gifu 501-1193, Japan}}

\author{K. Matsubara}
\affiliation{{Faculty of Engineering, Gifu University, 1-1 Yanagido, Gifu 501-1193, Japan}}

\author[0000-0002-8966-9856]{Y. Fukui}
\affiliation{{Faculty of Engineering, Gifu University, 1-1 Yanagido, Gifu 501-1193, Japan}}
\affiliation{{Department of Physics, Nagoya University, Furo-cho, Chikusa-ku, Nagoya 464-8601, Japan}}

\author{J. Nishi}
\affiliation{{Department of Physics and Astronomy, Graduate School of Science and Engineering, Kagoshima University, 1-21-35 Korimoto, Kagoshima, Kagoshima 890-0065, Japan}}

\author[0000-0001-9687-8237]{S. Einecke}
\affiliation{{School of Physical Sciences, The University of Adelaide, North Terrace, Adelaide, SA 5005, Australia}}

\author[0000-0002-4990-9288]{M. D. Filipovi\'{c}}
\affiliation{{Western Sydney University, Locked Bag 1797, Penrith South DC, NSW 1797, Australia}}

\author[0009-0001-3857-8671]{R. Kasai}
\affiliation{{Department of Physics and Astronomy, Graduate School of Science and Engineering, Kagoshima University, 1-21-35 Korimoto, Kagoshima, Kagoshima 890-0065, Japan}}

\author[0000-0001-5509-8218]{R. Matsusaka}
\affiliation{{Department of Physics and Astronomy, Graduate School of Science and Engineering, Kagoshima University, 1-21-35 Korimoto, Kagoshima, Kagoshima 890-0065, Japan}}
\affiliation{{Institute of Astronomy, Graduate School of Science, The University of Tokyo, 2-21-1 Osawa, Mitaka, Tokyo 181-0015, Japan}}

\author[0000-0002-9516-1581]{G. Rowell}
\affiliation{{School of Physical Sciences, The University of Adelaide, North Terrace, Adelaide, SA 5005, Australia}}

\author{H. Sodoh}
\affiliation{{National Institute of Technology, Sendai College 48 Nodayama, Medeshima-Shiote, Natori, Miyagi 981-1239, Japan}}

\author[0000-0002-8152-6172]{H. Suzuki}
\affiliation{{Faculty of Engineering, University of Miyazaki, Miyazaki 889-2192, Japan}}

\author[0009-0005-9445-7187]{Y. Shibata}
\affiliation{{Department of Physics, Nagoya University, Furo-cho, Chikusa-ku, Nagoya 464-8601, Japan}}
\affiliation{{Department of Physics and Astronomy, Graduate School of Science and Engineering, Kagoshima University, 1-21-35 Korimoto, Kagoshima, Kagoshima 890-0065, Japan}}

\author[0000-0002-2794-4840]{K. Tsuge}
\affiliation{{Faculty of Engineering, Gifu University, 1-1 Yanagido, Gifu 501-1193, Japan}}
\affiliation{{Institute for Advanced Study, Gifu University, 1-1 Yanagido, Gifu 501-1193, Japan}}

\author{H. Takaba}
\affiliation{{Faculty of Engineering, Gifu University, 1-1 Yanagido, Gifu 501-1193, Japan}}
\affiliation{{National Astronomical Observatory of Japan, Mitaka, Tokyo 181-8588, Japan}}

\author[0000-0002-0648-9826]{T. Handa}
\affiliation{{Department of Physics and Astronomy, Graduate School of Science and Engineering, Kagoshima University, 1-21-35 Korimoto, Kagoshima, Kagoshima 890-0065, Japan}}
\affiliation{{Amanogawa Galaxy Astronomy Research Center, Kagoshima University, 1-21-35 Korimoto, Kagoshima, Kagoshima 890-0065, Japan}}
\affiliation{{Division of Liberal Arts, Kogakuin University, 2665-1 Nakano-cho, Hachioji, Tokyo 192-0015, Japan}}

\begin{abstract}
We present a study of the interstellar medium associated with the two middle-aged supernova remnants (SNRs) W41 and G22.7–0.2, both detected in TeV gamma-rays. Using high-angular-resolution 
\co($J$~=~1–0) data from the Nobeyama 45-m telescope and \HI\ data from the VLA, we investigated the spatial and kinematic properties of molecular and atomic gas that interact with the SNRs. 
We identified associated clouds in the velocity ranges of +50–+80~\kms\ for W41 and +76–+110~\kms\ for G22.7–0.2.
Column density analysis indicates that target protons are dominated by molecular hydrogen, while atomic hydrogen contributes less than $\sim$10–15\% even after correction for self-absorption. 
The mean proton densities are $\sim$1.2$\times$10$^{3}$~\ccm\ for W41 and $\sim$5.3$\times$10$^{2}$~\ccm\ for G22.7–0.2. 
From the gamma-ray luminosities, we estimate the total energy of accelerated cosmic-ray protons as $W_\mathrm{p}$ $\sim$3$\times$10$^{47}$~erg for W41 and $\sim$1$\times$10$^{48}$~erg for G22.7–0.2, corresponding to 0.03–0.1\% of the canonical supernova explosion energy. 
These $W_\mathrm{p}$ values agree with the decreasing trend in $W_\mathrm{p}$ observed in the middle-aged SNRs within the previously reported SNR age–$W_\mathrm{p}$ relation.
\end{abstract}


\section{Introduction} \label{sec:intro}
Since their discovery by \cite{Hess1912}, the origin of Galactic cosmic rays, which consist primarily of relativistic protons with energies up to $\sim$3~PeV, has remained one of the long-standing problems in modern astrophysics. 
Supernova remnants (SNRs) are considered the most plausible accelerators of such cosmic-rays, because their shocks, propagating at velocities of $\sim$3,000–10,000~\kms, provide an ideal environment for diffusive shock acceleration \citep[DSA; e.g.,][]{bell1978MNRAS.182..147B,Blandford1978ApJ...221L..29B,Drury1983RPPh...46..973D}.
If SNRs are the dominant sources of Galactic cosmic rays, the total energy of accelerated cosmic-ray protons, $W_\mathrm{p}$, per supernova explosion can be estimated from the Galactic cosmic-ray energy density, the confinement time in the disk, and the supernova rate, yielding a conventional value of 10$^{49}$–10$^{50}$~erg \citep[see a review by][]{Gabici2013ASSP...34..221G}. 
However, this estimate has not yet been observationally confirmed.

To observationally validate this value, it is essential to study the interstellar medium (ISM) associated with hadronic gamma-ray SNRs. 
Cosmic-ray protons interact with ambient protons through p–p collisions to produce neutral pions, which subsequently decay into hadronic gamma rays. 
Thus, accurate determination of the total ISM proton content provides a direct means of deriving $W_\mathrm{p}$. 
In many previous observational and theoretical studies, the ISM was approximated by ionized gas with a uniform density, typically assumed to be $\sim$1~\ccm\ \citep[e.g.,][]{Ellison2010ApJ...712..287E,Lee2013ApJ...767...20L,Yang2014A&A...567A..23Y}.
However, recent radio observations have demonstrated that such assumptions are not justified. 
High-resolution data have revealed highly inhomogeneous ISM structures, including optically thick \HI, coincident with SNR shells, and have confirmed that the dominant target material consists of neutral molecular and atomic hydrogen \citep[e.g.,][]{Fukui2003IAUS..221P.224F,Fukui2012ApJ...746...82F,Fukui2015ApJ...798....6F,Fukui2017ApJ...850...71F,Yoshiike2013ApJ...768..179Y,aruga2022ApJ...938...94A,Sano2017ApJ...843...61S,Sano2018ApJ...867....7S,sano2021ApJ...923...15S,sano2021ApJ...919..123S,sano2022ApJ...933..157S}.

Even when the target gas density is derived by considering both molecular and atomic components and the $W_\mathrm{p}$ is estimated from the comparison with the observed gamma-ray luminosity, the physical interpretation of $W_\mathrm{p}$ requires careful attention. 
This is because gamma-ray emission from SNRs reflects the time evolution of cosmic-ray acceleration, diffusion, and escape.  
Recent studies have shown that cosmic-ray acceleration nearly terminates within a few hundred to a thousand years after the explosion, during the free-expansion phase, at which time $W_\mathrm{p}$ is determined \citep[e.g.,][]{Tsuji2021ApJ...907..117T,Suzuki2021ApJ...914..103S}.
\cite{sano2021ApJ...923...15S,sano2021ApJ...919..123S,sano2022ApJ...933..157S} analyzed 13 gamma-ray emitting SNRs older than $\sim$1~kyr where CR acceleration is nearly complete—and derived the relation between the SNR age and $W_\mathrm{p}$. 
They found a positive correlation for SNRs aged 1–6~kyr and a negative correlation for those older than 8~kyr, which they interpreted as a consequence of cosmic-ray diffusion and escape.
The increasing trend in young SNRs can be explained in the framework of age-limited acceleration \citep[e.g.,][]{Ohira2010A&A...513A..17O,Telezhinsky2012APh....35..300T}, where the finite age of the remnant constrains the maximum acceleration time. 
An alternative explanation is that the degree of cosmic-ray penetration into the stellar-wind shell evolves over time, resulting in a time-dependent gamma ray luminosity \citep[see also][]{Sano2023ApJ...958...53S}. 
In contrast, the decreasing trend observed in middle-aged SNRs is generally attributed to energy-dependent diffusion \citep[e.g.,][]{Aharonian1996A&A...309..917A,Gabici2013ASSP...34..221G}. 
In this process, high-energy cosmic rays gradually escape from the SNR shell and diffuse into the surrounding ISM, leading to a decrease in the inferred total proton energy $W_\mathrm{p}$ with increasing SNR age. 
Such behavior has also been discussed for W44 \citep[e.g.,][]{Uchiyama2012ApJ...749L..35U}.

However, the number of middle-aged SNRs with well-constrained $W_\mathrm{p}$ values remains limited. 
The next key step is to increase the statistical sample of such remnants and to observationally verify the unified picture of cosmic ray acceleration and diffusion throughout SNR evolution.

W41 and G22.7–0.2 are both middle-aged SNRs with detected gamma-ray emission, making them ideal laboratories for investigating the age–$W_\mathrm{p}$ relation.
Their ages are estimated to be $\sim$60~kyr and $\sim$20~kyr, respectively \citep[e.g.,][]{Stafford2019ApJ...884..113S,Suzuki2021ApJ...914..103S}. H.E.S.S. observations have identified HESS~J1834–087 and HESS~J1832–093 as their gamma-ray counterparts \citep[][]{HESS2015MNRAS.446.1163H,hess2018A&A...612A...1H}.
\cite{Castro2013ApJ...774...36C} performed broadband spectral modeling of Fermi-LAT and H.E.S.S. data for W41 and concluded that the gamma-ray emission is of hadronic origin. They argued that the leptonic scenario is disfavored, since reproducing the observed spectrum would require an electron-to-proton ratio as high as $\sim$10\%.
For G22.7–0.2, comparisons with Galactic ridge spectra suggest that its gamma-ray emission most likely originates from proton interactions with nearby molecular clouds, although its origin remains uncertain \citep[][]{HESS2015MNRAS.446.1163H,Tam2020ApJ...899...75T}. Interstellar molecular and atomic clouds associated with these SNRs have also been discussed in previous studies \citep[e.g.,][]{Leahy2008AJ....135..167L,frail2013ApJ...773L..19F,Su2014ApJ...796..122S,Su2015ApJ...811..134S}. Notably, \cite{frail2013ApJ...773L..19F} detected shock-excited 1720~MHz OH masers with a line-of-sight velocity of $\sim$+74~\kms\ toward the central radio shell of W41, providing strong evidence for interaction between the SNR shock and molecular clouds.
However, the limited angular resolution of previous CO data ($\sim$46\arcsec) has precluded detailed spatial and kinematic analyses of the associated gas, and the identification and quantitative characterization of the target ISM interacting with cosmic rays remain incomplete.

In this study, we aim to identify and analyze the associated ISM around W41 and G22.7–0.2 using high angular resolution CO and \HI\ data obtained with the Nobeyama 45-m Telescope and the VLA. 
The Nobeyama CO data provide an unprecedented angular resolution of $\sim$20\arcsec, enabling precise identification of molecular clouds interacting with the SNRs. The structure of this paper is as follows. 
Section \ref{sec:dataset} describes the observational data and analysis methods. Section \ref{sect_results} is divided into three subsections: Section \ref{sect_3_1} presents the distributions of radio continuum and gamma rays; Section \ref{sect_3_2} describes the spatial and velocity structures of the CO and \HI\ gas; and Section \ref{sect_3_3} discusses the \HI\ brightness dips at the locations of CO emission. The discussion and conclusions are given in Sections \ref{sect_discuss} and \ref{sect_conclusion}.

\section{Data sets} \label{sec:dataset}
\subsection{Radio continuum}
To investigate the spatial distribution of the SNR shells, we used the 20~cm radio continuum data taken from the Multi-Array Galactic Plane Imaging Survey \citep[MAGPIS;][]{Helfand2006AJ....131.2525H}. The spatial resolution is 6$\farcs$4 $\times$ 5$\farcs$4 with a position angle of 350\fdg0. The typical rms noise is $\sim$0.3~mJy~beam$^{-1}$.

\subsection{H{\scriptsize{I}}}
We used the \HI\ line data from the VLA Galactic Plane Survey \citep[VGPS;][]{stil2006AJ....132.1158S}.
The spatial resolution of \HI\ cube data was 60\arcsec.
The typical rms noise level is $\sim$2~K at a velocity channel width of 0.824~\kms.
Details on observing techniques and data processing are given in \cite{stil2006AJ....132.1158S}.

\subsection{H.E.S.S. TeV Gamma Rays}
We used archival gamma-ray Galactic Plane survey data obtained by the High Energy Stereoscopic System \citep[H.E.S.S.;][]{hess2018A&A...612A...1H}. In this study, we used the gamma-ray significance maps generated with a standard correlation radius $R_c$ = 0\fdg1.
The 68\% containment radii of point-spread function (PSF) is 0\fdg08. 
For more details, see \cite{hess2018A&A...612A...1H}.

\subsection{CO}
\subsubsection{Observations}

\co($J$~=~1–0) line data were obtained as part of the FOREST Unbiased Galactic plane Imaging Survey with the Nobeyama 45-m telescope \citep[FUGIN;][]{Umemoto2017PASJ...69...78U}.
These data were obtained using the multi-beam receiver, FOur-beam REceiver System on the 45-m Telescope \citep[FOREST;][]{Minamidani2016}.
The observations were made by the On-The-Fly (OTF) mapping mode \citep{Sawada2008PASJ...60..445S}.
The half-power beam width (HPBW) is $\sim$14\farcs4 at 115~GHz.
The overall map is made by connecting $1\degr \times 1\degr$ mosaic sub-maps.
The spectrometer was the Spectral Analysis Machine for the 45-m \citep[SAM45;][]{kuno2011} with 4,096 channels, providing a velocity resolution of 1.3~\kms\ at 115~GHz.
See \cite{Umemoto2017PASJ...69...78U} for a detailed observing strategy and OTF scan parameters.

\subsubsection{Data reduction}
Since published FUGIN data are contaminated with stripe structures along the scan direction, we performed self-reduction in \co($J$~=~1–0) line for the range of $21\fdg0 < l < 25\fdg0$, $|b| < 1\fdg0$.
Data reduction was performed using NOSTAR software provided by the Nobeyama Radio Observatory \citep[see Section 5.1 in][]{Sawada2008PASJ...60..445S}. For each $1\degr \times 1\degr$ mosaic sub-map, the following data processing was performed:
\begin{enumerate}
    \item Split raw data files into individual arrays.
    \item The intensity variation of each array was calibrated with daily mapping observations towards W51D [$\alpha_{\rm B1950} = 19^\mathrm{h}21^\mathrm{m}22\fs2, \delta_{\rm B1950} = 14\degr25\arcmin17\farcs0$].
    \item Baseline subtraction was performed with a third-order polynomial function. The baseline ranges are from $-$250 to $-$100 and from +200 to +350~\kms.
\end{enumerate}

Maps are produced as a cube FITS by gridding over the desired area.
The Gridding Convolution Function (GCF) was used the Bessel $\times$ Gaussian function.
Finally, we made a cude FITS with a spatial grid interval of 8\farcs5 and velocity interval of 0.65~\kms.
The effective angular resolution was 20\farcs2 for \co($J$~=~1–0). 
Additionally, a baseline subtraction was performed using a third-order polynomial function on the FITS of the produced cube. 
The baseline ranges were determined manually for each on-source scan.

The produced maps suffer from scanning noise (also known as \textit{scanning effect}) in addition to thermal noise.
Scanning effects are usually subtracted by the basket-weave method, which combines two maps produced from orthogonal scans \citep[e.g.,][]{Emerson1988A&A...190..353E,Muller2017A&A...606A..41M}.
However, in the $21\fdg0 < l < 25\fdg0$ region, only the scan parallel to the Galactic plane (X-scan) was observed due to instrumental trouble. In this study, we used the PRESS method \citep{Sofue1979A&AS...38..251S,Sofue2019PASJ...71..121S} for the subtraction of the scanning effects.

\section{Results}
\label{sect_results}
\subsection{Overview of Gamma-rays and Radio Continuum}
\label{sect_3_1}
\begin{figure*}[ht!]
\begin{center}
\includegraphics[width= 0.9\textwidth, trim={0 0 0 0}, clip]{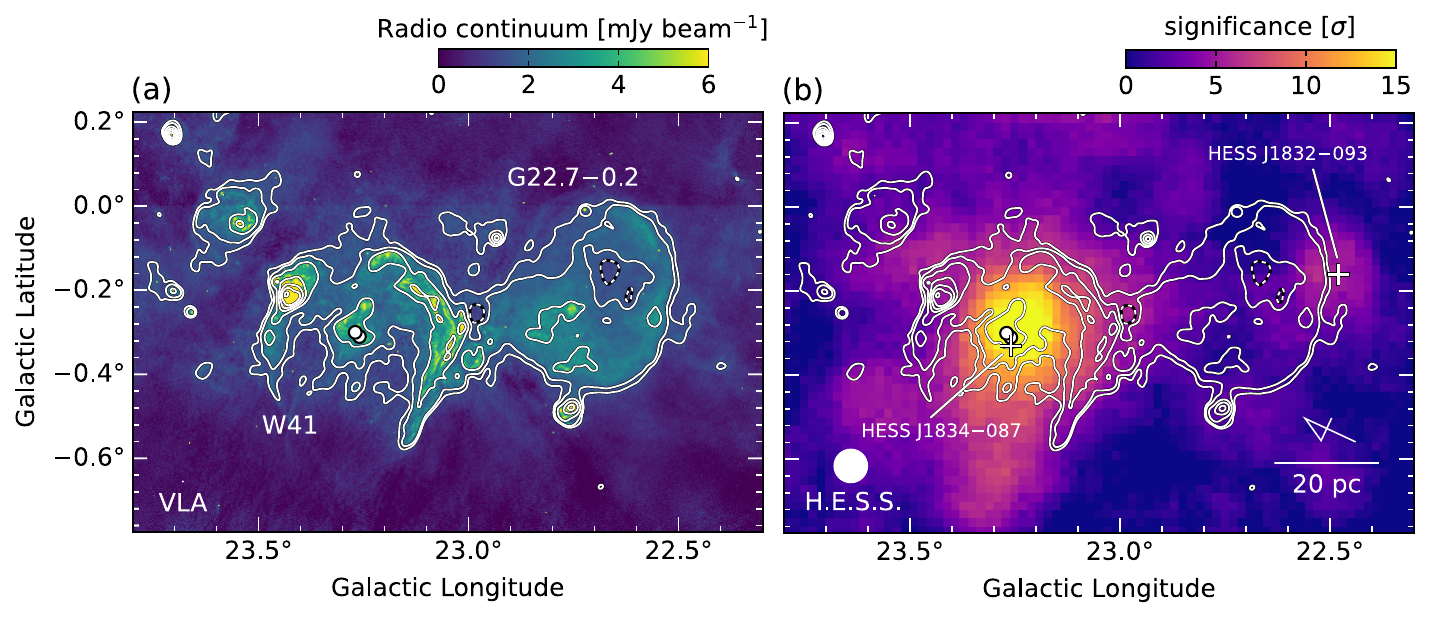}
\caption{Maps of (a) 20~cm radio continuum from the MAGPIS survey \citep{Helfand2006AJ....131.2525H}, (b) TeV gamma-ray significance map obtained from H.E.S.S. \citep{hess2018A&A...612A...1H} toward the SNRs W41 and G22.7–0.2. The superposed contours are 20~cm radio continuum. The contour levels are 1.40, 1.75, 2.80, 4.55, 7.00, 10.15, and 14.00~mJy~beam$^{-1}$. The circles indicates the positions of the 1720~MHz OH maser \citep{frail2013ApJ...773L..19F}. The positions of TeV gamma-ray sources are indicated by the white crosses in Figure \ref{fig:faceimage}(b).
\label{fig:faceimage}}
\end{center}
\end{figure*}

Figure \ref{fig:faceimage} shows the radio continuum and H.E.S.S. gamma-ray significance maps of W41 and G22.7–0.2.
Both SNRs exhibit radio shells with an angular diameter of $\sim$30$\arcmin$ and are located adjacent to each other around ($l$, $b$) = (23\fdg0, $-$0\fdg3). However, their morphologies and correspondence with the gamma-ray emission differ.

W41 shows a well-defined shell, with particularly bright emission at its center ($l$, $b$) = (23\fdg3, $-$0\fdg3) and along the southern rim ($l$, $b$) = (23\fdg0, $-$0\fdg3). In contrast, G22.7–0.2 has a relatively faint shell with a concave feature on its western side at ($l$, $b$) = (22\fdg8, $-$0\fdg1). In gamma-rays, HESS~J1834–087, related to W41, extends around the center of the shell with a peak significance of $\sim$17$\sigma$. HESS~J1832–093, related to G22.7–0.2, appears as a clumpy feature on southwest side of the shell at ($l$, $b$) = (22\fdg5, $-$0\fdg2), with a peak significance of $\sim$6$\sigma$.

Additionally, radio continuum brightest peaks on the two SNRs, ($l$, $b$) = (22\fdg7, $-$0\fdg5) and (23\fdg4, $-$0\fdg2). 
However, as Figure \ref{fig:faceimage} (b) shows, these do not coincide with the gamma-ray peaks. 
The former corresponds to the massive star-forming region G23.43–0.18 \citep[e.g.,][]{Leahy2008AJ....135..167L,Brunthaler2009ApJ...693..424B,Fujisawa2014PASJ...66...31F,Chibueze2025MNRAS.539..145C}, for which the distance has been estimated as 5.88$^{+1.37}_{-0.93}$~kpc based on annual parallax measurements of methanol masers \citep{Brunthaler2009ApJ...693..424B}. 
On the other hand, the kinematic distance of the molecular clouds related to W41 and G22.7–0.2 is $\sim$4.4~kpc \citep{frail2013ApJ...773L..19F,Su2014ApJ...796..122S,Su2015ApJ...811..134S}, suggesting a lack of physical association. 
The latter peak has been identified as either an SNR candidate \citep{Helfand2006AJ....131.2525H} or as an \HII\ region \citep{Thompson2006A&A...453.1003T}. 
Its spectral index has been measured as $-$0.98 \citep{Messineo2010ApJ...708.1241M}. 
In either case, these radio continuum sources are unrelated to the present study and are excluded from further discussion.

\subsection{Distribution of Radio continuum, CO and H\scriptsize{I}}
\label{sect_3_2}
\begin{figure*}[ht!]
\begin{center}
\includegraphics[width=0.90\textwidth, trim={230 480 210 330}, clip]{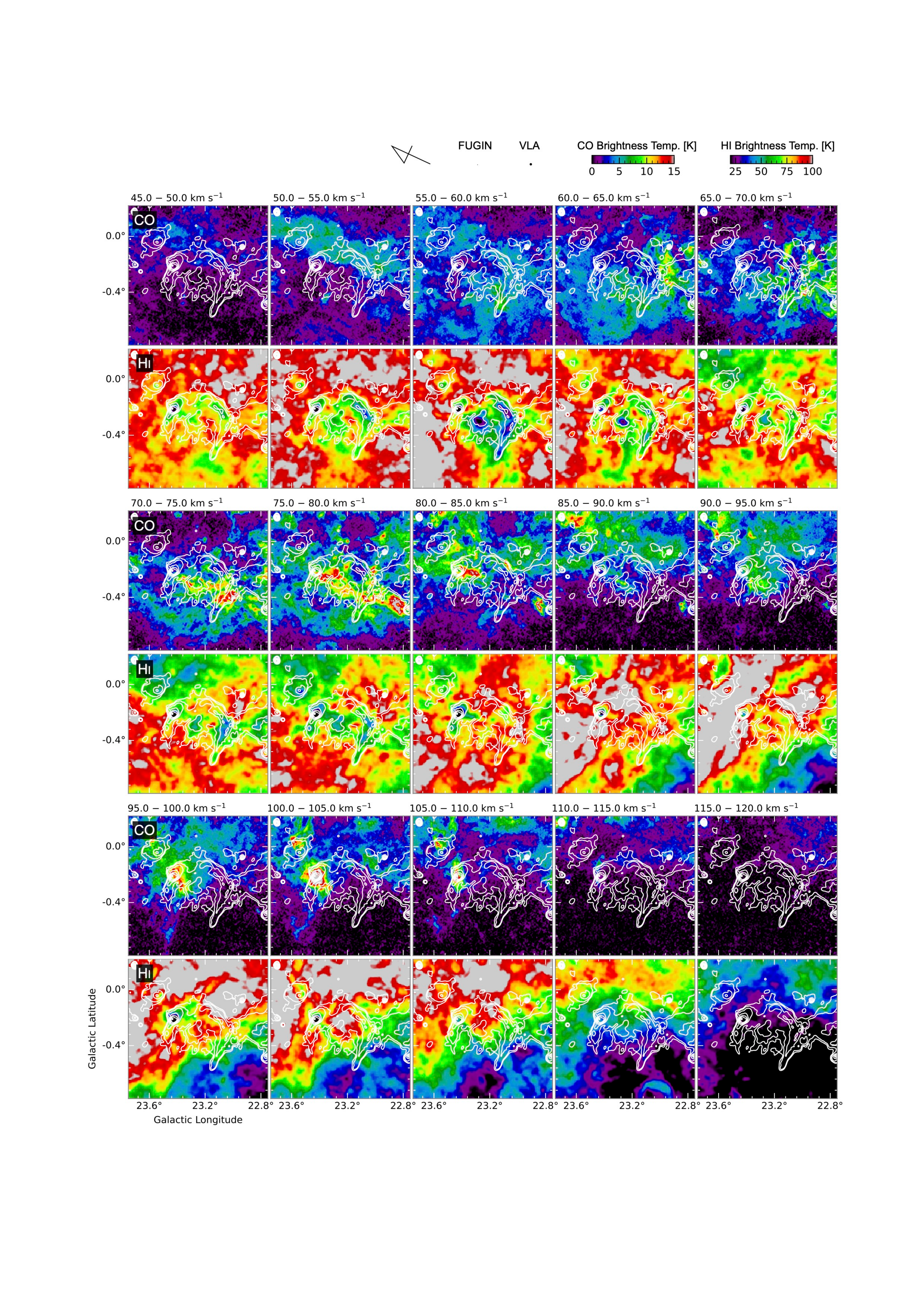}
\caption{Velocity channel distributions of the \co($J$~=~1–0) and \HI\ brightness temperatures towards W41 superposed with the same radio continuum contours as in Figure \ref{fig:faceimage}(a). Each panel shows intensity distributions averaged every 5~\kms\ in a velocity range from +45 to +120~\kms. The color bar is shown on top of the set of panels.
\label{fig:channel_map_CO_1}}
\end{center}
\end{figure*}

\begin{figure*}[ht!]
\begin{center}
\includegraphics[width=0.88\textwidth, trim={230 590 210 245}, clip]{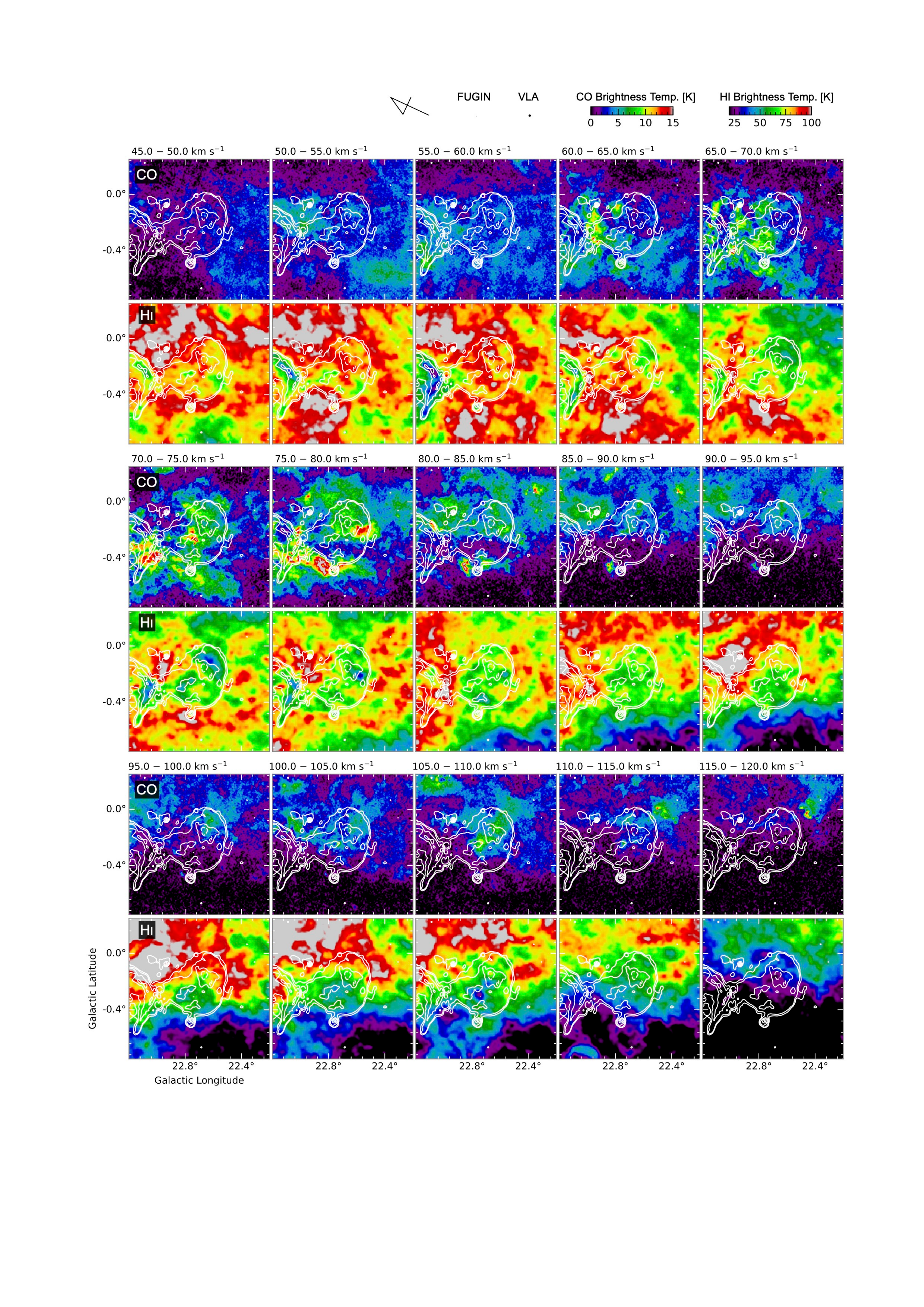}
\caption{Same velocity channel maps as shown in Figure \ref{fig:channel_map_CO_1}, but it is centered on G22.7–0.2.
\label{fig:channel_map_CO_2}}
\end{center}
\end{figure*}

Figures \ref{fig:channel_map_CO_1} and \ref{fig:channel_map_CO_2} present the \co($J$~=~1–0) and \HI\ channel maps toward W41 and G22.7–0.2, respectively. In this paper, we focus on the velocity range from +45 to +120~\kms, which includes the line-of-sight velocity of the shock-excited 1720~MHz OH maser.
The \HI channel maps presented in Figures \ref{fig:channel_map_CO_1} and \ref{fig:channel_map_CO_2} use a limited color scale optimized to emphasize the large-scale spatial correspondence with the radio continuum shell. 
It should be noted that they do not display the full dynamic range, including the absorption features.

For W41, CO emission in the +55–+60~\kms\ range surrounds the radio shell from the southern to western sides and continues to the northern boundary. 
A diffuse, weak component with a brightness temperature of $\sim$5~K is also present at the shell center. 
At +70–+80~\kms, molecular gas associated with GMC G23.3–0.4 extends across the shell, showing a bright CO peak of $\sim$11~K at ($l$, $b$) = (23\fdg3, $-$0\fdg3).
This CO emission corresponds well with a weak \HI\ feature ($\sim$60~K in brightness temperature) seen in the same velocity range. 
In \HI\ maps, weak clouds with brightness temperatures of 30–60~K spatially coincide with the radio shell at $\sim$+60~\kms. 
Such correspondence is widely found over the +50–+80~\kms\ range. 
In addition, the northern side of the shell at ($l$, $b$) = (23\fdg4, $-$0\fdg2) shows an \HI\ depression across nearly the entire velocity range (see Figure \ref{fig:spectral_analysis}, position A, for the detailed \HI\ line profile).
Clumpy CO features are also seen in this region, corresponding to the massive star-forming region G23.43–0.18 \citep[e.g.,][]{Leahy2008AJ....135..167L,Brunthaler2009ApJ...693..424B,Fujisawa2014PASJ...66...31F,Chibueze2025MNRAS.539..145C}.

For G22.7–0.2, CO emission spreads diffusely within the radio shell over +50–+115~\kms.
Notably, at +75–+80~\kms, a CO cloud at the eastern edge ($l$, $b$) = (22\fdg8, $-$0\fdg4) shows well spatial agreement with the radio shell. 
A CO clump with $\sim$12~K brightness temperature is also identified in the southern shell at ($l$, $b$) = (22\fdg6, $-$0\fdg1). 
In comparison with \HI, no clear correlation is seen in the +45–+70~\kms\ range. However, at +70–+95~\kms\ and +105–+110~\kms, weak \HI\ clouds with brightness temperatures of $\sim$60~K align well with the shell. 
Furthermore, local depressions of $\sim$30~K are observed at ($l$, $b$) = (22\fdg6, $-$0\fdg1) in +70–+75~\kms, at ($l$, $b$) = (22\fdg6, $-$0\fdg2) in +75–+80~\kms, and at ($l$, $b$) = (22\fdg8, $-$0\fdg3) in +105–+110~\kms. These depressions spatially coincide with CO emission.

\begin{figure*}[ht!]
\begin{center}
\includegraphics[width=0.8\textwidth, trim={0 72 70 0}, clip]{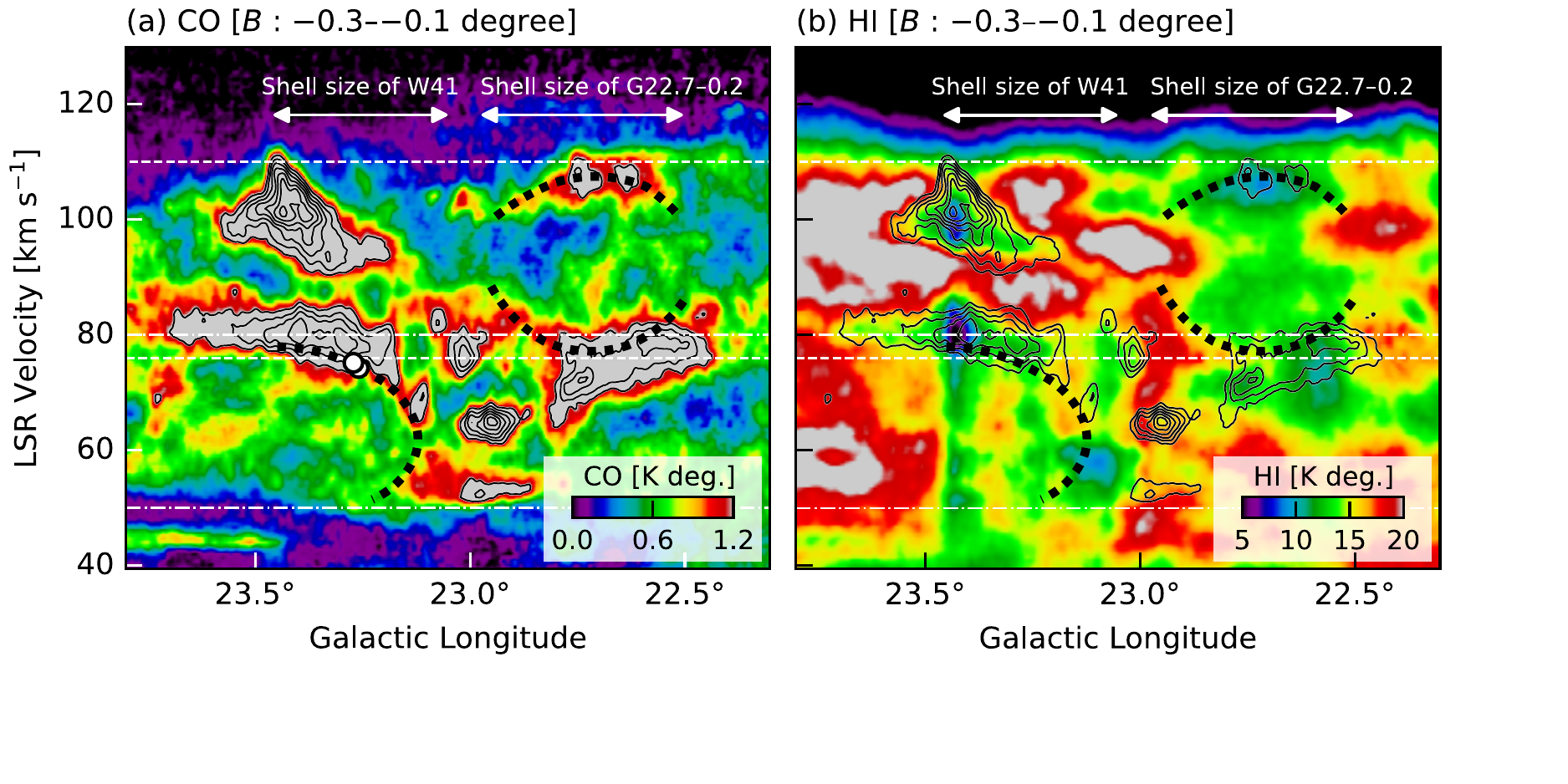}
\caption{Longitude–Velocity diagrams of (a) \co($J$~=~1–0) and (b) \HI. The integrated range is from $-$0\fdg3 to $-$0\fdg1 in the Galactic Latitude. The double-headed arrows indicate the diameter of each SNR shell. The dashed black curves show the expanding gas motion for each SNR (see the text). Horizontal dashed and one-dot chain lines indicate the integration velocity ranges for each SNR. The superposed contours in each panel indicate \co($J$~=~1–0), whose contour levels are 1.2, 1.5, 1.8, 2.1, 2.4, 2.7, 3.0, and 3.3, K~degree. The white filled circles represent the positions of the 1720~MHz OH masers \citep{frail2013ApJ...773L..19F}.
\label{fig:vl_maps}}
\end{center}

\end{figure*}

Figure \ref{fig:vl_maps} shows the longitude–velocity maps of CO and \HI. Two cavity-like structures are identified in the CO map: one at 23\fdg1 $< l <$ 23\fdg5 and +50–+80~\kms, and another at 22\fdg5 $< l <$ 23\fdg0 and +75–+110~\kms.
The longitudinal extents of these structures are nearly correspond to the diameters of their respective SNR shells. 
In the corresponding regions, the \HI\ maps also exhibit faint emission ($\sim$12~K~deg). 
Figure \ref{fig:vl_maps}(b) shows that low-brightness \HI\ features of $\sim$10–13~K~deg. are observed at ($l$, $v$) = (22\fdg6, +80~\kms), (22\fdg7, +110~\kms), (23\fdg4, +80~\kms), and (23\fdg4, +100~\kms), which is in good agreement with the CO distribution. 
Conversely, at ($l$, $v$) = (23\fdg1, +60~\kms), faint \HI\ ($\sim$10~K~deg.) is detected without corresponding CO emission. 
Additionally, aligned low-brightness \HI\ features are seen over +40–+110~\kms\ around $l$ = 23\fdg4, with pronounced depressions at +80 and +100~\kms, which again coincide with CO. 
In contrast, no corresponding CO emission is detected in the +40–+70~\kms\ range. 
This series of faint \HI\ features corresponds to the massive star-forming region G23.43–0.18, whih is located at $v_\mathrm{LSR} \sim$+100~\kms.

\subsection{Analysis of the CO and H{\scriptsize{I}} spectral}
\label{sect_3_3}
\begin{figure*}[ht!]
\begin{center}
\includegraphics[width=\linewidth, trim={0 0 0 0}, clip]{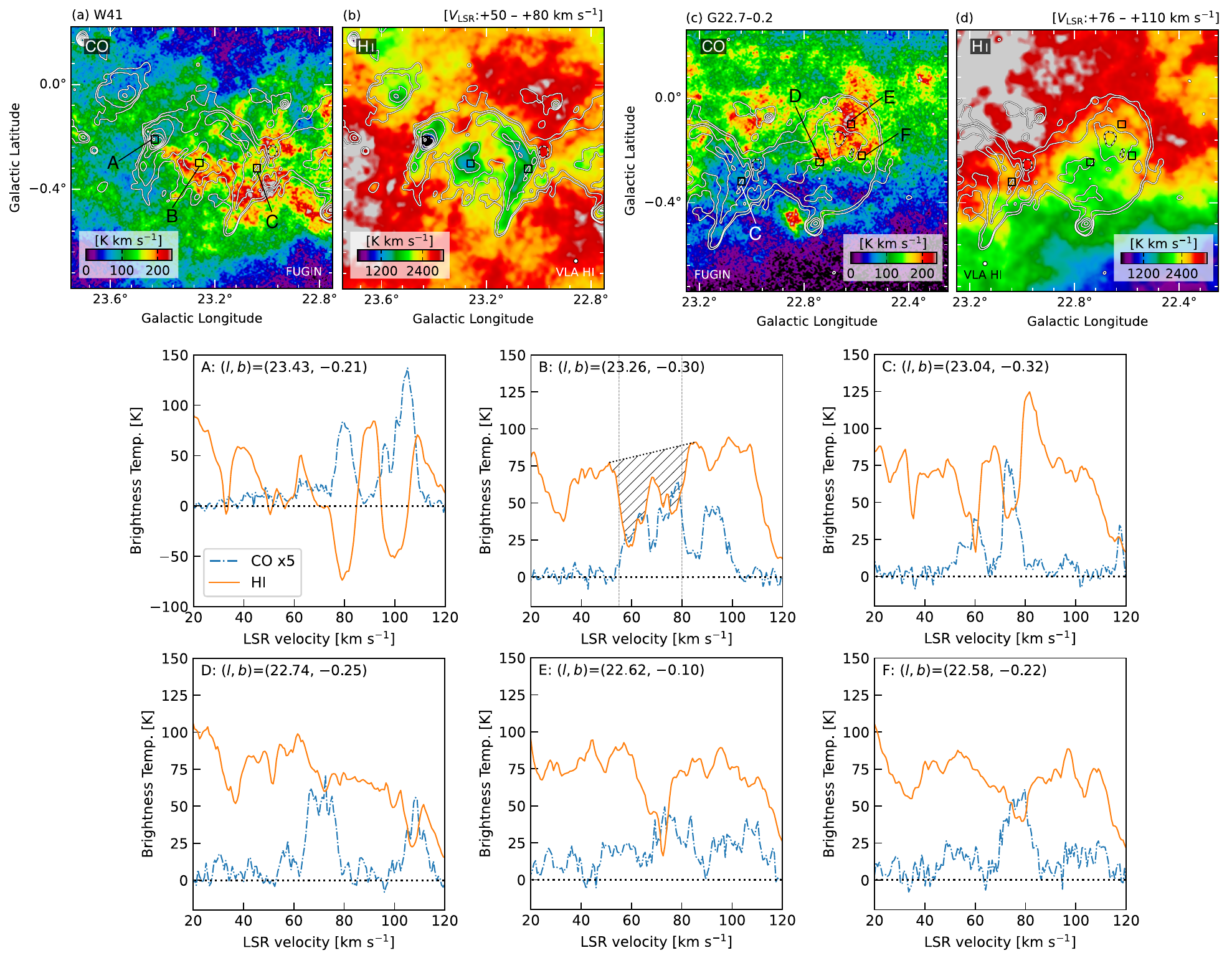}
\caption{Integrated intensity maps of the \co($J$~=~1–0) and \HI\ emission toward (a), (b) W41 in the velocity range $v_\mathrm{LSR}$ = +50–+80~\kms, and (c), (d) G22.7–0.2 in the range $v_\mathrm{LSR}$ = +76–+110~\kms. Black boxes mark the positions where line profiles were extracted. Line profiles shown below indicate \co($J$~=~1–0) (Blue) and \HI\ (orange) at six positions: A ($l$, $b$) = (23\fdg43, $-$0\fdg21), B (23\fdg26, $-$0\fdg30), C (23\fdg04, $-$0\fdg32), D (22\fdg74, $-$0\fdg25), E (22\fdg62, $-$0\fdg10), and F (22\fdg58, $-$0\fdg22). The shaded area and vertical dashed lines in Position B indicates an expected background \HI\ profile and a velocity range from +55 to +80~\kms.
\label{fig:spectral_analysis}}
\end{center}
\end{figure*}

To distinguish whether the low-brightness \HI\ coincident with the radio continuum are due to absorption by the background continuum or self-absorption caused by cold \HI\ embedded in molecular gas, we compared the \HI\ and CO spectra. 
For both W41 and G22.7–0.2, we selected regions in which the low \HI\ brightness temperature corresponds to the radio shell. 
Then, we constructed line profiles at each position. 
Figure \ref{fig:spectral_analysis} shows the integrated intensity maps of CO and \HI, together with representative line profiles at selected positions. 
Of the positions shown in Figures \ref{fig:spectral_analysis}(a)–(d), positions A–C correspond to W41 and positions D–F correspond to G22.7–0.2.

At positions A–C, the \HI\ emission generally peaks near +70~\kms and +85~\kms. 
Position A lies in a region containing a massive star-forming region in the background, where deep \HI\ dips reaching the lowest brightness ($\sim$$-90$~K) are observed between +80 and +100~\kms. 
These features are consistent with absorption against strong background continuum emission \citep[brightness temperature $\sim$180~K;][]{stil2006AJ....132.1158S}. 
However, CO emission coincident with the \HI\ dips is also detected at both velocities. 
This suggests that both absorption against the continuum background and self-absorption by cold \HI\ clouds located in front of it may both contribute. 
In contrast, at positions B and C, prominent \HI\ dips with depths of 30–50~K and line widths of a few \kms\ are observed near +60~\kms\ and +78~\kms, showing well spatial and velocity correspondence with \co\ emission.

For positions D–F, the features are less prominent. 
However, \HI\ peaks are seen around +90~\kms\ and +110~\kms, showing slight saturation. 
Additionally, \HI\ dips with depths of 20–50~K are found near +75~\kms\ and +100~\kms\ at all positions, again showing good correspondence with CO emission. 

To interpret these behaviors, we briefly summarize the \HI–H$_2$ transition. 
It is well recognized that atomic hydrogen is converted into molecular hydrogen on dust grain surfaces as the column density and ultraviolet shielding increase \citep[e.g.,][]{Allen1977ApJ...212..396A}. 
Accordingly, the distributions of cold \HI\ and CO emission are generally similar, as both trace the dense, shielded regions within molecular clouds. 
Based on these results, we conclude that the \HI\ absorption dips observed at positions B–F are predominantly due to self-absorption by cold \HI\ within the CO-emitting molecular clouds.

\section{Discussion}
\label{sect_discuss}
\subsection{Molecular clouds associated with the SNR W41 and G22.7–0.2}
\label{sect_4_1}
Previous ISM studies of W41 and G22.7–0.2 have suggested the presence of associated molecular clouds, based on the detection of a shock-excited 1720~MHz OH maser at $\sim$+74~\kms\ and molecular clouds crossing both SNRs \citep[e.g.,][]{Lee2013ApJ...767...20L,frail2013ApJ...773L..19F,Su2014ApJ...796..122S,Su2015ApJ...811..134S,hogge2019ApJ...887...79H}.
However, comprehensive analyses using both CO and \HI\ data are lacking, and interacting clouds have not been firmly identified.
In this section, we argue that the molecular and atomic clouds physically associated with W41 and G22.7–0.2 are found in the velocity ranges of +50–+80~\kms\ and +76–+110~\kms, respectively.
Identifying the SNR-associated clouds requires not only investigating spatial distributions but also considering velocity structures and \HI\ absorption features.
We then evaluate the consistency of these results with previously reported gas components. 
Specifically, Section \ref{sect_4_1_1} describes the spatial distribution of the clouds and \HI\ absorption, Section \ref{sect_4_1_2} addresses their velocity structures, and Section \ref{sect_4_1_3} discusses consistency with previous studies.

\subsubsection{Spatial Distributions and H{\scriptsize{I}} absorption}
First, we show the spatial correspondence between the radio continuum shells and the molecular gas (Figures \ref{fig:channel_map_CO_1}, \ref{fig:channel_map_CO_2}, and Section \ref{sect_3_2}).
In W41, CO emission in the +55–+60~\kms\ range is distributed so as to wrap around the southern to western to northern boundaries of the shell. 
In addition, at +70–+80~\kms, a bright CO feature with $T_\mathrm{mb}\sim$11~K is found at the shell center ($l$, $b$) = (23\fdg3, $-$0\fdg3), coinciding in both position and velocity with the shock-excited 1720~MHz OH maser. 
These features strongly suggest interactions between the SNR shock and molecular clouds. 
In G22.7–0.2, CO emission includes diffuse components extending across the shell interior as well as concentrations along its periphery. 
Notably, at +75–+80~\kms, a CO cloud near ($l$, $b$) = (22\fdg8, $-$0\fdg4) shows excellent spatial agreement with the radio shell, forming a typical structure indicative of shock-cloud interaction. 
This correspondence implies that magnetic field amplification at the shocked cloud surface enhances synchrotron emission \citep[e.g.,][]{Inoue2009ApJ...695..825I,Inoue2012ApJ...744...71I,Sano2013ApJ...778...59S}.

The \HI\ distribution also shows noteworthy spatial correlations. 
In W41, weak \HI\ is detected at +55–+60~\kms\ closely trace the radio shell, and in several regions, weak \HI\ coincides with CO features. 
In G22.7–0.2, weak \HI\ is found both inside the shell and along its southern rim, overlapping spatially with CO. 
These intensity depressions are not caused by absorption against the radio continuum background, but rather by self-absorption from cold \HI\ embedded within molecular clouds (see Section \ref{sect_3_3}). 
Thus, the observed \HI\ characteristics indicate that the atomic and molecular gas locate the same three-dimensional regions, supporting the conclusion that both are physically associated with the SNRs.
\label{sect_4_1_1}

\subsubsection{Expanding Gas Motion} 
\label{sect_4_1_2}
Figure \ref{fig:vl_maps} shows cavity-like velocity structures in the $l$–$v$ diagrams, which provide compelling evidence of interactions between the SNR shocks and the surrounding medium. 
Such cavity features show expanding gas motions and are typically interpreted as wind-blown bubbles formed by stellar winds. 
These structures are thought to result from the combined action of strong winds from the progenitor star and the subsequent supernova shock wave \citep[e.g.,][]{Koo1990ApJ...364..178K,Koo1991ApJ...382..204K,Koo1992ApJ...388...93K,Dwarkadas2005ApJ...630..892D}. 

We found cavity-like structures extending over +50 to +80~\kms\ toward W41 and +75 to +110~\kms\ toward G22.7–0.2 (Figure \ref{fig:vl_maps}). 
From the $l$–$v$ diagrams, we defined the systemic velocity as the velocity where the cavity shows its maximum spatial extent and derived the expansion velocities accordingly.
For W41, the systemic velocity is approximately +63~\kms\ with an expansion velocity of $\sim$17~\kms ; for G22.7–0.2, the corresponding values are +92.5~\kms\ and $\sim$17.5~\kms.
%The estimated expansion velocities are $\pm$15~\kms\ around +65~\kms\ for W41 and $\pm$17.5~\kms\ around +92.5~\kms\ for G22.7–0.2.
Although these expanding velocities are smaller than typical SN shock velocities, the apparent bubble sizes are roughly consistent with the observed radio continuum shells (Figure \ref{fig:vl_maps}). 
This suggests that the SN shocks propagate rapidly through the bubble interiors because the gas density and its gradient are extremely low, reducing deceleration \citep[e.g.,][]{Weaver1977ApJ...218..377W,Dwarkadas2005ApJ...630..892D,Broersen2014MNRAS.441.3040B}.
Therefore, the detected bubbles are likely to have been formed by the strong stellar winds from the progenitor and are now interacting with supernova shock waves.

This scenario is further supported by the clear spatial correspondence between the shock-excited 1720~MHz OH maser and the CO clouds in W41. 
The OH maser has a velocity of $\sim$+74~\kms\ and is located near the center of the radio shell (see Figure \ref{fig:faceimage}(a)).
In velocity space, the maser lies on the boundary of the CO cloud, within the extent of the bubble (see Figure \ref{fig:vl_maps}(a)). 
Although the projection geometry differs, such associations between molecular clouds and OH masers have also been reported in other Galactic SNRs \citep[e.g.,][]{Arikawa1999PASJ...51L...7A,sano2021ApJ...923...15S}, providing a representative example of SNR shocks interacting with molecular clouds.

The expansion velocities identified in this study are larger than the values of several Galactic and Magellanic SNRs \citep[e.g., a few \kms\ to $\sim$10~\kms\ ;][]{Landecker1989MNRAS.237..277L,Fukui2012ApJ...746...82F,Sano2017ApJ...843...61S,Sano2018ApJ...867....7S,Sano2019ApJ...881...85S,sano2022ApJ...933..157S,Kuriki2018ApJ...864..161K,Fukushima2020ApJ...897...62F,aruga2022ApJ...938...94A,Yeung2023PASJ...75..384Y,Ito2025ApJ...978..123I}. 
This suggests that the observed motions may not only reflect the intrinsic evolution of the SNRs but also be influenced by their surrounding environment. 
The GMC G23.3–0.4, to which both W41 and G22.7–0.2 belong, hosts multiple SNRs, young massive stars, and red supergiants.
Spectroscopic studies have reported that massive star formation has been ongoing in this region over several tens of Myr \citep{Messineo2014A&A...569A..20M}.
It is therefore probable that the progenitor winds of the currently observed SNRs began expanding within low-density environments created by previous generations.
This possibility is further supported by the nearly circular and symmetric shapes of the radio shells of both SNRs (Figure \ref{fig:faceimage}(a)), which are consistent with their expansion into a relatively homogeneous and low-density medium. 
Nevertheless, a detailed investigation of this issue lies beyond the scope of the present study. 
To better understand the connection between the surrounding environment and the subsequent evolution of SNRs, future observational and theoretical studies will be required.

\subsubsection{Consistency and Comparison with Previous Studies}
\label{sect_4_1_3}
Based on the above results, we conclude that the molecular and atomic clouds physically associated with W41 and G22.7–0.2 correspond to the velocity ranges of +50–+80~\kms\ and +76–+110~\kms, respectively. 
This conclusion is supported by their spatial distributions, velocity structures, and the presence of OH masers. 
The wind-blown bubbles identified in both SNRs overlap in the +76–+80~\kms\ range, suggesting a possible physical interaction with GMC G23.3–0.4, which is observed in the +70–+80~\kms\ velocity range. 
This interpretation is consistent with previous CO and \HI\ studies \citep[e.g.,][]{Albert2006ApJ...643L..53A,Leahy2008AJ....135..167L,frail2013ApJ...773L..19F,Su2014ApJ...796..122S,Su2015ApJ...811..134S}.

Our analysis reveals that the two SNRs exhibit different systemic velocities. 
These results provide clues to the three-dimensional spatial relation between the SNRs and their associated molecular clouds. 
The GMC G23.3–0.4 shows a continuous spatial distribution extending from the central region of the W41 radio shell to the eastern rim of the G22.7–0.2 shell (Figure \ref{fig:spectral_analysis}(a)), making it natural to regard them as part of the same molecular cloud. 
As shown in Figure \ref{fig:vl_maps}(a), the interacting components of the SNRs relative to the GMC indicate that W41 lies on the redshifted side with respect to the systemic velocity of bubble, while G22.7–0.2 lies on the blueshifted side. 
From the spatial correspondence between the radio shells and the molecular clouds, W41 shows interaction with the molecular gas at the shell center, whereas G22.7–0.2 shows interaction along the shell rim (Figures \ref{fig:channel_map_CO_1}, \ref{fig:channel_map_CO_2}, and Section \ref{sect_4_1_2}). These results suggest that the three-dimensional configuration places W41 in front of the GMC and G22.7–0.2 slightly behind or at nearly the same distance as the GMC. 

\begin{figure}[t!]
\begin{center}
\includegraphics[width=0.75\linewidth, trim={0 0 0 0}, clip]{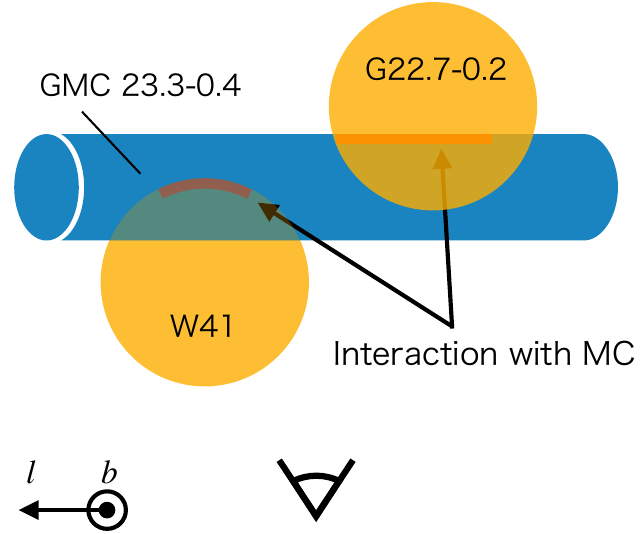}
\caption{Schematic top-down view of the relative positions of the two SNRs and the GMC. The orange ellipses represent the SNR bubbles and the blue cylinder shows the GMC~G23.3–0.4. The thick red lines highlight the regions where the SNRs interact with the GMC. The shell of W41 is located in front of the GMC, while G22.7–0.2 is located above the GMC along the Galactic latitude direction (see also Figure \ref{fig:channel_map_CO_2}).
\label{fig:3d_location}}
\end{center}
\end{figure}

Based on these considerations, we present in Figure \ref{fig:3d_location}, a schematic view of the line-of-sight locations of the two SNRs and the GMC. This schematic is consistent with the spatial and kinematic analyses described in this section. 
In any case, determining the true three-dimensional locations of the objects solely from radio continuum and line data remains extremely challenging. 
Future measurements of stellar three-dimensional motions in this region will provide critical clues for clarifying the positional relationship between the molecular clouds and the SNRs. 
Observations with Gaia and the upcoming infrared astrometric satellite JASMINE \citep[e.g.,][]{Gouda2012ASPC..458..417G} are particularly promising in this regard.

\subsection{Total ISM Protons}
\label{sect_4_2}
From Section \ref{sect_4_1}, the velocity ranges of the clouds associated with W41 and G22.7–0.2 have been identified. 
In this section, we estimate the total number of ISM protons in both molecular and atomic phases in order to compare with the gamma-ray emission. 
If the gamma rays are of hadronic origin, their spatial distribution should correspond to that of the target ISM. 
To this end, we first identify the clouds spatially associated with the gamma-ray-emitting regions in each SNR (Section \ref{sect_4_2_1}). 
We then estimate the molecular gas mass from the CO data (Section \ref{sect_4_2_2}), followed by an evaluation of the atomic gas mass from the \HI\ data (Section \ref{sect_4_2_3}).

\subsubsection{Identification of Target Clouds}
\label{sect_4_2_1}

\begin{figure*}[ht!]
\begin{center}
\includegraphics[width=0.9\textwidth, trim={0 0 0 0}, clip]{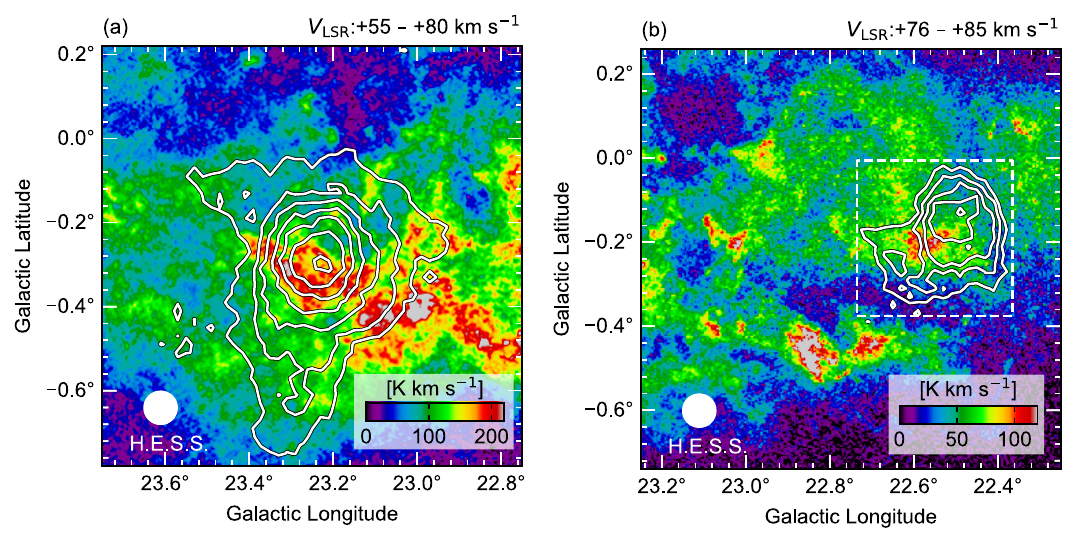}
\caption{Integrated intensity maps of the \co($J$~=~1–0) emission toward (a) W41 at $V_\mathrm{LSR}$ = +55–+80~\kms and (b) G22.7–0.2 at $V_\mathrm{LSR}$ = +76–+85~\kms. 
The superposed contours are TeV gamma-ray significance maps.
The contour levels are 5, 7, 9, 11, 13, and 15 $\sigma$ for (a) and 2, 3, 4, and 5 $\sigma$ for (b). The dashed rectangle in (b) shows the region of the TeV gamma-ray source HESS J1832–093. The PSF of H.E.S.S is indicated by the gray circle shown in the lower left corner.
\label{fig:comp_co_gamma}}
\end{center}
\end{figure*}

According to Figure \ref{fig:faceimage}(b), W41 shows a gamma-ray peak at the central region of its radio shell ($l$, $b$) = (23\fdg3,$-$0\fdg3), while G22.7–0.2 exhibits a peak on the southern part of its shell ($l$, $b$) = (22\fdg5,$-$0\fdg2). 
Thus, the ISM clouds corresponding to these locations are expected to serve as the target gas. 
On the other hand, because the \HI\ spectra exhibit self-absorption due to cold \HI\ (see Section \ref{sect_3_3}), it is difficult to establish a direct spatial correlation with the gamma rays. Therefore, we used the \co($J$~=~1–0) data to identify the relevant clouds. 

Analysis of the channel maps (Figures \ref{fig:channel_map_CO_1} and \ref{fig:channel_map_CO_2}) shows that molecular clouds in the velocity range of +55–+80~\kms\ for W41 and +76–+85~\kms\ for G22.7–0.2 exhibit well spatial correspondence with the gamma-ray emission. 
Figure \ref{fig:comp_co_gamma} presents the CO integrated intensity maps over these velocity intervals, overlaid with the gamma-ray significance contours. 
In both SNRs, strong CO emission is found to coincide with the gamma-ray peaks. 
We therefore identify these velocity ranges as the most plausible candidates for the ISM gas associated with the gamma-ray emission.
It should be noted that the angular resolution of the gamma-ray data ($\sim$0\fdg08) is lower than that of the CO and \HI\ data; consequently, our comparison is limited to integrated spatial distributions rather than a direct point-to-point correspondence.

\subsubsection{Molecular Gas}
\label{sect_4_2_2}
The column density of molecular hydrogen, $N(\mathrm{H}_2)$, was derived from the integrated intensity of the \co($J$~=~1–0) emission using the CO–H$_2$ conversion factor:
\begin{equation}
N(\mathrm{H}_2) = X_{\rm CO} \, W(\mathrm{CO}) \ \ (\mathrm{cm}^{-2}),
\end{equation}
where $X_{\rm CO}$ is the CO--H$_2$ conversion factor and $W(\mathrm{CO})$ is the integrated intensity of the \co($J$~=~1–0) line. In this study, we adopted $X_{\rm CO} = 1.0 \times 10^{20} \ \mathrm{cm^{-2} \ ((K \ km \ s^{-1})^{-1}})$ \citep{Okamoto2017ApJ...838..132O}. 
Using this method, the peak molecular column densities of the associated clouds in both SNRs were estimated to be $\sim$(1–2)$\times 10^{22}$~\scm. 
The proton column density of the molecular component is defined as $N_{\rm p}(\mathrm{H}_2) = 2 \times N(\mathrm{H}_2)$.

The molecular cloud mass, $M$, was then calculated as
\begin{equation}
M = m_{\rm H} \, \mu \, \Omega D^2 \sum_i N_i(\mathrm{H}_2),
\end{equation}
where $m_{\rm H}$ is the mass of a hydrogen atom, $\mu = 2.8$ is the mean molecular weight per hydrogen molecule, $\Omega$ is the solid angle per pixel, $D = 4.4$ kpc is the distance to the SNRs, and $N_i(\mathrm{H}_2)$ is the molecular column density at pixel $i$. 
The cloud size was defined as the half-maximum contour of the CO integrated intensity. 
Based on this procedure, we obtained molecular cloud masses of $\sim 7.2 \times 10^{4} \ M_\odot$ for W41 and $\sim 2.4 \times 10^{4} \ M_\odot$ for G22.7–0.2.

\subsubsection{Atomic Gas}
\label{sect_4_2_3}
In general, the atomic hydrogen column density is estimated under the assumption that the \HI\ emission is optically thin.
If this assumption holds, the column density of interstellar protons, $N_\mathrm{p}(\mathrm{H}{\textsc{i}})$, can be expressed as \citep{Dickey1990ARA&A..28..215D}:
\begin{equation}
N_{\rm p}(\mathrm{H}{\textsc{i}}) = 1.823 \times 10^{18} \int T_{\rm b}(v) , dV \ \ (\mathrm{cm}^{-2}),
\end{equation}
where $T_{\rm b}$ is the observed \HI\ brightness temperature and $v$ is the radial velocity.
However, as discussed in Section \ref{sect_3_3}, the \HI\ spectra show absorption dips coincident with the CO velocity ranges, indicating that self-absorption by cold \HI\ embedded in molecular clouds cannot be neglected.
Therefore, corrections for optical depth effects must be considered.
Indeed, \citep{Fukui2012ApJ...746...82F}, based on an analysis of RX J1713.7–3946, showed that the optical depth of self-absorbing \HI\ is typically $\tau \sim 0.7$, resulting in an increase of $N_{\rm p}(\mathrm{H}{\textsc{i}})$ by a factor of $\sim$1.2 compared to the optically thin assumption.

The observed \HI\ brightness temperature as a function of radial velocity, $T_\mathrm{L}(v)$, can be expressed by the radiative transfer equation (e.g., \citealt{Sato1978AJ.....83.1607S}):
\begin{eqnarray}
T_\mathrm{L}(v) = T_\mathrm{s} \left[1 - e^{-\tau(v)} \right] +\left[ T_\mathrm{L}^{\mathrm{BG}}(v) 
+ T_\mathrm{C}^{\mathrm{BG}} \right] e^{-\tau(v)} \nonumber \\
-\left( T_\mathrm{C}^{\mathrm{FG}} + T_\mathrm{C}^{\mathrm{BG}} \right),
\end{eqnarray}
where $T_\mathrm{L}(v)$ is the observed \HI\ brightness temperature, $T_\mathrm{s}$ is the spin temperature, $\tau(v)$ is the optical depth of the cold \HI\ cloud, $T_\mathrm{L}^{\mathrm{FG}}(v)$ and $T_\mathrm{L}^{\mathrm{BG}}(v)$ represent the foreground and background \HI\ emission, and $T_\mathrm{C}^{\mathrm{FG}}$ and $T_\mathrm{C}^{\mathrm{BG}}$ denote the brightness temperatures of the foreground and background radio continuum emission, respectively.
The background \HI\ component, $T_\mathrm{L}^{\mathrm{BG}}(v)$, was estimated using the method described by \citet{Sato1978AJ.....83.1607S}, in which a linear interpolation across the \HI\ dip is applied (see the dashed line at position B in Figure \ref{fig:spectral_analysis}). This approach generally provides a lower limit to the true absorption. Even when a parabolic interpolation is adopted instead of a linear one, the difference in the resulting correction remains within $\sim$30\% \citep{Sato1978AJ.....83.1607S}. 
In the present study, we adopt the linear interpolation. 
Furthermore, based on the radio continuum emission measured from Figure \ref{fig:faceimage}(a), we assume $T_\mathrm{C}^{\mathrm{FG}}=0$ when estimating the optical depth.

We estimated the column density of cold \HI, $N_\mathrm{p}(\mathrm{H}{\textsc{i}})$, toward position B in Figure \ref{fig:spectral_analysis}, where the spatial correspondence between the gamma-ray emission and molecular clouds is strong and a clear \HI\ dip is observed. 
Distinct \HI\ dips are seen at $+60$~\kms\ and +80~\kms. The column density of cold \HI, taking into account optical depth effects, can be derived as
\begin{equation}
N_\mathrm{p}(\mathrm{H}{\textsc{i}}) = 1.823 \times 10^{18} \int T_\mathrm{s}\tau(v)dV \ \ \mathrm{(cm^{-2})},
\end{equation}
where $T_\mathrm{s}$ denotes the spin temperature. 
Direct measurements of the spin temperature of cold \HI\ are generally difficult; however, for cold \HI\ embedded in molecular clouds, $T_\mathrm{s}$ is expected to be comparable to the kinematic temperature of the molecular gas. 
This is because the typical density of molecular clouds ($\sim10^{3}$~\ccm) is high enough for collisions with H$_2$ molecules to thermalize the \HI\ with the molecular gas. 
Since the critical density for collisional excitation of the \co($J$~=~1–0) transition is also $\sim10^{3}$~\ccm, this assumption is reasonable in regions where CO emission is detected. 
From the depth of the \HI\ dips at position B, we estimate $T_\mathrm{s}\sim20$~K for the +60~\kms\ component and $\lesssim50$~K for the +80~\kms\ component. 
The CO peak temperature suggests a molecular gas temperature of $\sim$15~K. Based on these considerations, we calculated $N_{\rm p}(\mathrm{H}{\textsc{i}})$ for $T_\mathrm{s}$~=~20, 30, and 40~K within the velocity range of +55–+80~\kms\ (as defined in Section \ref{sect_4_2_1}). 
The resulting \HI\ column densities are (4.5–5.5)$\times10^{21}$~\scm, with maximum optical depths of $\tau\sim$0.8–1.1. 
These values are approximately a factor of 2.5 higher than those derived under the optically thin assumption.
At position B, the molecular proton column density is $N_\mathrm{p}(\mathrm{H_2}) \sim 3.7 \times 10^{22}$~\scm; therefore, even after correcting for self-absorption, the contribution of \HI\ remains limited to 12–15\% of the total proton column density. 
Applying the same method to position F (toward G22.7–0.2), we obtained $N_\mathrm{p}(\mathrm{H}{\textsc{i}}) \sim (0.8$–$1.0)\times10^{21}$~\scm \ with maximum optical depths of $\tau$$\sim$0.4–0.5, corresponding to only 6–8\% of the total proton column density. 
These results indicate that the bulk of the interstellar protons associated with the gamma-ray emission originates from molecular hydrogen rather than atomic hydrogen.

From the above analysis, we conclude that the dominant component of the target protons for the gamma-ray emission in both W41 and G22.7–0.2 is molecular hydrogen. 
Finally, using the molecular cloud masses derived in Section \ref{sect_4_2_2}, we estimated the mean proton number density. 
The cloud volume was calculated from the equivalent radius based on the projected area of each molecular cloud. 
As a result, we obtained $n_\mathrm{p} \sim 1.2 \times 10^{3}$~\ccm\ for W41 and $n_\mathrm{p} \sim 5.3 \times 10^{2}$~\ccm\ for G22.7–0.2.

\subsection{Total Energy of CR Protons}
\label{sect_4_3}

\begin{figure*}[ht!]
\begin{center}
\includegraphics[width=0.57\textwidth, trim={48 40 50 35}, clip]{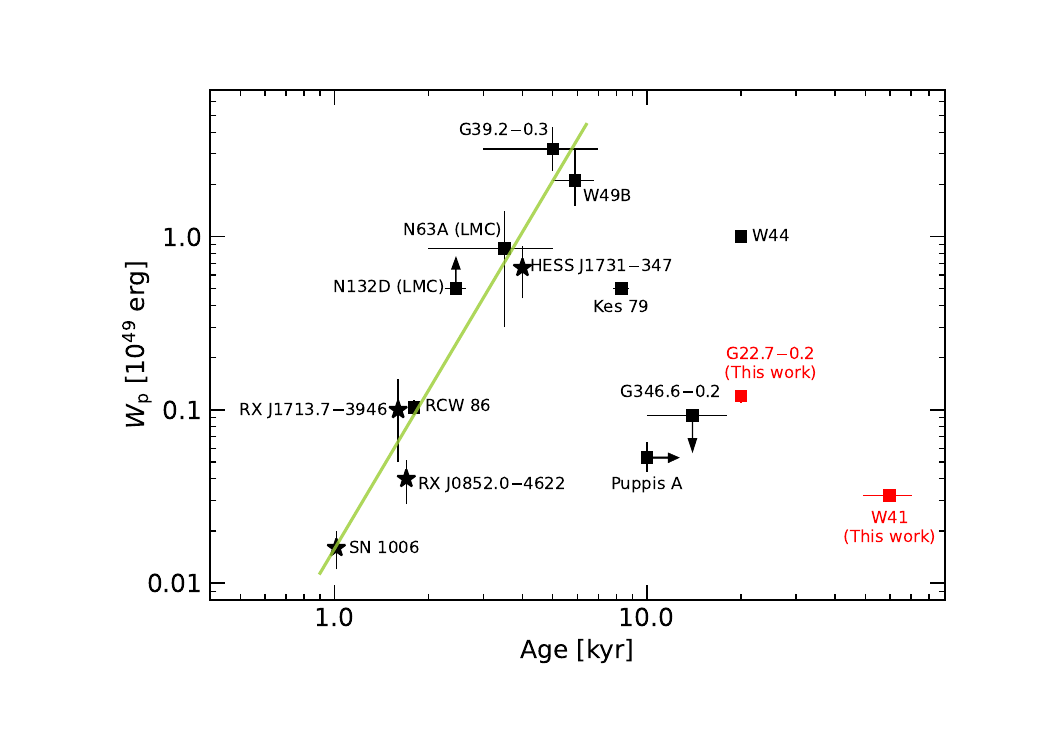}
\caption{Scatter plot between the SNR age and the total energy of the cosmic-ray protons $W_\mathrm{p}$ \citep[e.g.,][]{sano2021ApJ...923...15S,sano2021ApJ...919..123S,sano2022ApJ...933..157S,aruga2022ApJ...938...94A}. The green line indicates the linear regression of the double-logarithmic plot applying least-squares fitting to the data points with the ages of SNRs below 6~kyr. The plots of SN~1006, RX~J0852.0–4622, RX~J1713.7–3946 and HESS~J1731–347 consider the contribution of hadronic gamma-rays to the total gamma-rays \citep{Fukuda_2014,fukui2021ApJ...915...84F,fukui2024ApJ...961..162F,sano2022ApJ...933..157S}.
\label{fig:age_wp_plot}}
\end{center}
\end{figure*}

In this section, we estimate the total energy of accelerated cosmic-ray protons ($W_\mathrm{p}$) in both SNRs and discuss the relation between SNR age and $W_\mathrm{p}$. 
Previous studies have reported that for young SNRs with ages $\lesssim$~6~kyr, $W_\mathrm{p}$ tends to increase with age, while for middle-aged SNRs with ages $\gtrsim$~8~kyr, $W_\mathrm{p}$ decreases \citep[e.g.,][]{sano2021ApJ...923...15S,sano2021ApJ...919..123S,aruga2022ApJ...938...94A}.
If this observational trend holds, both W41 and G22.7–0.2, which are classified as middle-aged SNRs, are expected to exhibit lower $W_\mathrm{p}$ values compared to the peak.
In this study, we focus in particular on the diffusion of cosmic rays in middle-aged SNRs.

The total energy of accelerated CR protons, $W_\mathrm{p}$, can be derived under the assumption that the hadronic process dominates, as follows \citep[e.g.,][]{Aharonian2006A&A...449..223A}:
\begin{equation}
    W_\mathrm{p} \sim t_{pp \rightarrow \pi^0} \, L_\gamma \ \ \mathrm{(erg)},
\end{equation}
where $t_{pp \rightarrow \pi^0} \sim 4.5 \times 10^{13} \left( \frac{n_\mathrm{p}}{100 \ {\rm cm}^{-3}} \right)^{-1} \ {\rm s}$
is the cooling time of CR protons through inelastic p–p interactions, and $L_\gamma$ is the gamma-ray luminosity in units of erg~s$^{-1}$.
From \cite{Suzuki2022ApJ...924...45S}, based on GeV and TeV spectra obtained from Fermi-LAT and H.E.S.S., the gamma-ray luminosity associated with W41 is estimated to be $L_\gamma \sim 8.27(7.74\textsc{–}8.89) \times 10^{34} \ {\rm erg \ s^{-1}}$.
Using $n_\mathrm{p} \sim 1.2 \times 10^{3}$~\ccm, we obtain $W_\mathrm{p} \sim (3.2 \pm 0.2) \times 10^{47} \ {\rm erg}$.
In the case of G22.7–0.2, the gamma-ray luminosity is estimated to be $L_\gamma \sim 1.36 \,(1.23\textsc{–}1.51) \times 10^{35} \, {\rm erg \, s^{-1}}$, based on pion-decay model fitting to the combined Fermi and H.E.S.S. spectra assuming a spectral index of 2.6 \citep[][]{HESS2015MNRAS.446.1163H}. Adopting $n_\mathrm{p} \sim$ 5.3 $\times 10^{2}$~\ccm\ yields $W_\mathrm{p} \sim (1.2 \pm 0.1) \times 10^{48} \, {\rm erg}$.
The $W_\mathrm{p}$ values obtained in this study correspond to only 0.03–0.1\% of the typical kinematic energy released by a supernova explosion ($\sim$10$^{51}$~erg).

Figure \ref{fig:age_wp_plot} shows SNR age–$W_\mathrm{p}$ relation for 15 gamma-ray SNRs, including W41 and G22.7–0.2. 
Both remnants show values consistent with other middle-aged SNRs such as Puppis A \citep{aruga2022ApJ...938...94A}, and G346.6–0.2 \citep{sano2021ApJ...923...15S}.
Our results support the interpretation that a large fraction of CR protons escape from the SNR shell on timescales of $\sim$10$^{4}$~yr, leading to the reduced $W_\mathrm{p}$ observed at present. 
This scenario is also supported by estimates of the CR diffusion length \citep[see also][]{aruga2022ApJ...938...94A}.
An important point is that the $W_\mathrm{p}$ of W41 is comparable to that of SN~1006, the youngest SNR ($\sim$1.0~kyr) in the previously age–$W_\mathrm{p}$ plot. 
This provides an important constraint on the timescale of acceleration and escape of CRs over $\sim$1–60~kyr. 
It should be noted, however, that $W_\mathrm{p}$ for most SNRs is derived solely from spectral energy distribution modeling. 
This method has inherent uncertainties by a factor of 2–3 due to the difficulty in disentangling hadronic from leptonic gamma-ray components \citep[][]{Inoue2012ApJ...744...71I}. 
In addition, the estimate of $n_\mathrm{p}$ involves uncertainties associated with the CO-to-H$_2$ conversion factor, typically at the level of $\sim 30\%$. Nevertheless, these uncertainties do not alter the overall trend observed in the age–$W_\mathrm{p}$ relation.

\section{Conclusions}
\label{sect_conclusion}
In this study, we investigated the interaction between interstellar gas and two middle-aged gamma-ray SNRs, W41 and G22.7–0.2, using high-resolution CO and \HI \ data obtained with the Nobeyama 45-m telescope and the VLA. 
We identified the velocity ranges of molecular and atomic clouds physically associated with the two SNRs as +50–+80~\kms \ for W41 and +76–+110~\kms \ for G22.7–0.2. 
These clouds show good spatial correspondence with the radio shells, and in the case of W41, also with shock-excited OH masers. 
The cavity-like structures observed in the $l$–$v$ diagrams are interpreted as low-density bubbles created by stellar winds that are now interacting with the subsequent supernova shocks.

From column density analyses of the molecular and atomic gas, we found that molecular hydrogen dominates the total ISM proton content associated with both SNRs. 
The contribution from atomic hydrogen is less than 10\%, and this remains the case even after correcting for self-absorption. 
The average proton densities were estimated to be $\sim 1.2 \times 10^{3}$~\ccm \ for W41 and $\sim 5.3 \times 10^{2}$~\ccm \ for G22.7–0.2.

By comparing with the gamma-ray luminosities, we estimated the total energy of accelerated CR protons to be $W_\mathrm{p} \sim 3.2 \times 10^{47}$~erg for W41 and $\sim$ 1.2 $\times 10^{48}$~erg for G22.7–0.2.
These values correspond to only 0.03–0.1\% of the canonical kinematic energy of a supernova explosion ($\sim 10^{51}$~erg). Additionally, these values are consistent with the reported age–$W_\mathrm{p}$ relation in previous studies. 
In particular, the $W_\mathrm{p}$ of W41 is comparable to that of SN~1006, a younger SNR ($\sim$1.0~kyr), providing important constraints on the evolutionary timescale of CR proton acceleration and escape.

Future prospects include studying the interaction of escaped CRs with surrounding molecular clouds, which can produce secondary hadronic gamma-ray emission. 
Recent gamma-ray observations have reported a number of unidentified sources \citep[e.g.,][]{Aharonian2008A&A...477..353A,Abeysekara2017Sci...358..911A}, some of which may be explained by such secondary emission from escaped CRs. 
Indeed, observational evidence for this scenario has already been suggested in SNRs such as W28 and W44 \citep[e.g.,][]{Aharonian2008A&A...481..401A,Uchiyama2012ApJ...749L..35U}. 
For W41 and G22.7–0.2, molecular clouds other than those identified in this study could also serve as potential targets for escaped CRs. 
Future high-sensitivity gamma-ray observations with CTA will be crucial for detecting such secondary hadronic gamma-ray emission. 
Exploring the interaction between escaped CRs and the ISM will provide valuable constraints on the spatial and energy dependence of CR diffusion, thereby advancing our understanding of the underlying diffusion mechanisms.

\begin{acknowledgments}
We are grateful to the anonymous referee for carefully reading our manuscript and giving us thoughtful suggestions, which improved our paper. The 45-m radio telescope is operated by the Nobeyama Radio Observatory, a branch of the National Astronomical Observatory of Japan (NAOJ). This publication makes use of data from FUGIN, FOREST Unbiased Galactic plane Imaging survey with the Nobeyama 45-m telescope, a legacy project in the Nobeyama 45-m radio telescope. The scientific results reported in this article are based on data obtained from the H.E.S.S. Data Archive. We express our sincere gratitude to all the staff who contributed to the construction and operation of H.E.S.S. We thank Mr. Zachary Smeaton for his helpful discussions and valuable comments on the manuscript. This work was supported by a University Research Support Grant from NAOJ. It was also supported by JSPS KAKENHI Grant Numbers JP24H00246 (HS). This work was also supported by CCI Holdings Co., Ltd. This manuscript was compiled using Overleaf, a very useful free online LaTeX editor.
\end{acknowledgments}


\vspace{5mm}
\facilities{Nobeyama 45-m radio telescope (NRO45), 
            The Very Large Array (VLA), 
            High Energy Stereoscopic System (H.E.S.S).}
            
\software{astropy \citep{astropy_2013,astropy_2018}, 
          aplpy \citep{aplpy2012,aplpy2019},
          matplotlib \citep{Hunter_2007},
          numpy \citep{harris_2020},
          Scipy \citep{Virtanen2020SciPy-NMeth}.
          }


%\software{astropy \citep{2013A&A...558A..33A,2018AJ....156..123A},  
%          Cloudy \citep{2013RMxAA..49..137F}, 
%          Source Extractor \citep{1996A&AS..117..393B}
%          }



%\appendix


\bibliography{references}{}

@ARTICLE{Suzuki2021ApJ...914..103S,
       author = {{Suzuki}, Hiromasa and {Bamba}, Aya and {Shibata}, Shinpei},
        title = "{Quantitative Age Estimation of Supernova Remnants and Associated Pulsars}",
      journal = {\apj},
         year = 2021,
        month = jun,
       volume = {914},
       number = {2},
          eid = {103},
        pages = {103},
          doi = {10.3847/1538-4357/abfb02},
archivePrefix = {arXiv},
       eprint = {2104.10052},
 primaryClass = {astro-ph.HE},
       adsurl = {https://ui.adsabs.harvard.edu/abs/2021ApJ...914..103S}
}

@ARTICLE{Allen1977ApJ...212..396A,
       author = {{Allen}, M. and {Robinson}, G.~W.},
        title = "{The molecular composition of dense interstellar clouds.}",
      journal = {\apj},
         year = 1977,
        month = mar,
       volume = {212},
        pages = {396-415},
          doi = {10.1086/155059},
       adsurl = {https://ui.adsabs.harvard.edu/abs/1977ApJ...212..396A}
}

@ARTICLE{Sano2023ApJ...958...53S,
       author = {{Sano}, H. and {Yamane}, Y. and {van Loon}, J. Th. and {Furuya}, K. and {Fukui}, Y. and {Alsaberi}, R.~Z.~E. and {Bamba}, A. and {Enokiya}, R. and {Filipovi{\'c}}, M.~D. and {Indebetouw}, R. and {Inoue}, T. and {Kawamura}, A. and {Laki{\'c}evi{\'c}}, M. and {Law}, C.~J. and {Mizuno}, N. and {Murase}, T. and {Onishi}, T. and {Park}, S. and {Plucinsky}, P.~P. and {Rho}, J. and {Richards}, A.~M.~S. and {Rowell}, G. and {Sasaki}, M. and {Seok}, J. and {Sharda}, P. and {Staveley-Smith}, L. and {Suzuki}, H. and {Temim}, T. and {Tokuda}, K. and {Tsuge}, K. and {Tachihara}, K.},
        title = "{ALMA Observations of Supernova Remnant N49 in the Large Magellanic Cloud. II. Non-LTE Analysis of Shock-heated Molecular Clouds}",
      journal = {\apj},
         year = 2023,
        month = nov,
       volume = {958},
       number = {1},
          eid = {53},
        pages = {53},
          doi = {10.3847/1538-4357/acffbe},
archivePrefix = {arXiv},
       eprint = {2311.02180},
 primaryClass = {astro-ph.GA},
       adsurl = {https://ui.adsabs.harvard.edu/abs/2023ApJ...958...53S}
}

@ARTICLE{Messineo2014A&A...569A..20M,
       author = {{Messineo}, Maria and {Menten}, Karl M. and {Figer}, Donald F. and {Davies}, Ben and {Clark}, J. Simon and {Ivanov}, Valentin D. and {Kudritzki}, Rolf-Peter and {Rich}, R. Michael and {MacKenty}, John W. and {Trombley}, Christine},
        title = "{Massive stars in the giant molecular cloud G23.3-0.3 and W41}",
      journal = {\aap},
         year = 2014,
        month = sep,
       volume = {569},
          eid = {A20},
        pages = {A20},
          doi = {10.1051/0004-6361/201322822},
archivePrefix = {arXiv},
       eprint = {1408.3558},
 primaryClass = {astro-ph.GA},
       adsurl = {https://ui.adsabs.harvard.edu/abs/2014A&A...569A..20M}
}

@ARTICLE{Sano2013ApJ...778...59S,
       author = {{Sano}, H. and {Tanaka}, T. and {Torii}, K. and {Fukuda}, T. and {Yoshiike}, S. and {Sato}, J. and {Horachi}, H. and {Kuwahara}, T. and {Hayakawa}, T. and {Matsumoto}, H. and {Inoue}, T. and {Yamazaki}, R. and {Inutsuka}, S. and {Kawamura}, A. and {Tachihara}, K. and {Yamamoto}, H. and {Okuda}, T. and {Mizuno}, N. and {Onishi}, T. and {Mizuno}, A. and {Fukui}, Y.},
        title = "{Non-thermal X-Rays and Interstellar Gas Toward the {\ensuremath{\gamma}}-Ray Supernova Remnant RX J1713.7-3946: Evidence for X-Ray Enhancement around CO and H I Clumps}",
      journal = {\apj},
         year = 2013,
        month = nov,
       volume = {778},
       number = {1},
          eid = {59},
        pages = {59},
          doi = {10.1088/0004-637X/778/1/59},
archivePrefix = {arXiv},
       eprint = {1304.7722},
 primaryClass = {astro-ph.GA},
       adsurl = {https://ui.adsabs.harvard.edu/abs/2013ApJ...778...59S}
}

@ARTICLE{Virtanen2020SciPy-NMeth,
  author  = {Virtanen, Pauli and Gommers, Ralf and Oliphant, Travis E. and
            Haberland, Matt and Reddy, Tyler and Cournapeau, David and
            Burovski, Evgeni and Peterson, Pearu and Weckesser, Warren and
            Bright, Jonathan and {van der Walt}, St{\'e}fan J. and
            Brett, Matthew and Wilson, Joshua and Millman, K. Jarrod and
            Mayorov, Nikolay and Nelson, Andrew R. J. and Jones, Eric and
            Kern, Robert and Larson, Eric and Carey, C J and
            Polat, {\.I}lhan and Feng, Yu and Moore, Eric W. and
            {VanderPlas}, Jake and Laxalde, Denis and Perktold, Josef and
            Cimrman, Robert and Henriksen, Ian and Quintero, E. A. and
            Harris, Charles R. and Archibald, Anne M. and
            Ribeiro, Ant{\^o}nio H. and Pedregosa, Fabian and
            {van Mulbregt}, Paul and {SciPy 1.0 Contributors}},
  title   = {{{SciPy} 1.0: Fundamental Algorithms for Scientific
            Computing in Python}},
  journal = {Nature Methods},
  year    = {2020},
  volume  = {17},
  pages   = {261--272},
  adsurl  = {https://rdcu.be/b08Wh},
  doi     = {10.1038/s41592-019-0686-2},
}

@ARTICLE{Sano2019ApJ...881...85S,
       author = {{Sano}, H. and {Matsumura}, H. and {Yamane}, Y. and {Maggi}, P. and {Fujii}, K. and {Tsuge}, K. and {Tokuda}, K. and {Alsaberi}, R.~Z.~E. and {Filipovi{\'c}}, M.~D. and {Maxted}, N. and {Rowell}, G. and {Uchida}, H. and {Tanaka}, T. and {Muraoka}, K. and {Takekoshi}, T. and {Onishi}, T. and {Kawamura}, A. and {Minamidani}, T. and {Mizuno}, N. and {Yamamoto}, H. and {Tachihara}, K. and {Inoue}, T. and {Inutsuka}, S. and {Voisin}, F. and {Tothill}, N.~F.~H. and {Sasaki}, M. and {McClure-Griffiths}, N.~M. and {Fukui}, Y.},
        title = "{Discovery of Shocked Molecular Clouds Associated with the Shell-type Supernova Remnant RX J0046.5-7308 in the Small Magellanic Cloud}",
      journal = {\apj},
         year = 2019,
        month = aug,
       volume = {881},
       number = {1},
          eid = {85},
        pages = {85},
          doi = {10.3847/1538-4357/ab2ade},
archivePrefix = {arXiv},
       eprint = {1904.04836},
 primaryClass = {astro-ph.GA},
       adsurl = {https://ui.adsabs.harvard.edu/abs/2019ApJ...881...85S}
}

@ARTICLE{Sano2018ApJ...867....7S,
       author = {{Sano}, H. and {Yamane}, Y. and {Tokuda}, K. and {Fujii}, K. and {Tsuge}, K. and {Nagaya}, T. and {Yoshiike}, S. and {Filipovi{\'c}}, M.~D. and {Alsaberi}, R.~Z.~E. and {Barnes}, L. and {Onishi}, T. and {Kawamura}, A. and {Minamidani}, T. and {Mizuno}, N. and {Yamamoto}, H. and {Tachihara}, K. and {Maxted}, N. and {Voisin}, F. and {Rowell}, G. and {Yamaguchi}, H. and {Fukui}, Y.},
        title = "{Molecular Clouds Associated with the Type Ia SNR N103B in the Large Magellanic Cloud}",
      journal = {\apj},
         year = 2018,
        month = nov,
       volume = {867},
       number = {1},
          eid = {7},
        pages = {7},
          doi = {10.3847/1538-4357/aae07c},
archivePrefix = {arXiv},
       eprint = {1806.10299},
 primaryClass = {astro-ph.GA},
       adsurl = {https://ui.adsabs.harvard.edu/abs/2018ApJ...867....7S}
}

@ARTICLE{Ito2025ApJ...978..123I,
       author = {{Ito}, D. and {Sano}, H. and {Nakazawa}, K. and {Mitsuishi}, I. and {Fukui}, Y. and {Sudou}, H. and {Takaba}, H.},
        title = "{CO Observations of the Type Ia Supernova Remnant 3C 397 by the Nobeyama 45 m Radio Telescope: Possible Evidence for the Single-degenerated Explosion}",
      journal = {\apj},
         year = 2025,
        month = jan,
       volume = {978},
       number = {2},
          eid = {123},
        pages = {123},
          doi = {10.3847/1538-4357/ad95f5},
archivePrefix = {arXiv},
       eprint = {2405.11285},
 primaryClass = {astro-ph.HE},
       adsurl = {https://ui.adsabs.harvard.edu/abs/2025ApJ...978..123I}
}

@ARTICLE{Yeung2023PASJ...75..384Y,
       author = {{Yeung}, Paul K.~H. and {Bamba}, Aya and {Sano}, Hidetoshi},
        title = "{Multiwavelength studies of G298.6-0.0: An old GeV supernova remnant interacting with molecular clouds}",
      journal = {\pasj},
         year = 2023,
        month = apr,
       volume = {75},
       number = {2},
        pages = {384-396},
          doi = {10.1093/pasj/psad006},
archivePrefix = {arXiv},
       eprint = {2212.01851},
 primaryClass = {astro-ph.HE},
       adsurl = {https://ui.adsabs.harvard.edu/abs/2023PASJ...75..384Y}
}

@ARTICLE{Fukushima2020ApJ...897...62F,
       author = {{Fukushima}, Kotaro and {Yamaguchi}, Hiroya and {Slane}, Patrick O. and {Park}, Sangwook and {Katsuda}, Satoru and {Sano}, Hidetoshi and {Lopez}, Laura A. and {Plucinsky}, Paul P. and {Kobayashi}, Shogo B. and {Matsushita}, Kyoko},
        title = "{Element Stratification in the Middle-aged SN Ia Remnant G344.7-0.1}",
      journal = {\apj},
         year = 2020,
        month = jul,
       volume = {897},
       number = {1},
          eid = {62},
        pages = {62},
          doi = {10.3847/1538-4357/ab94a6},
archivePrefix = {arXiv},
       eprint = {2005.09664},
 primaryClass = {astro-ph.HE},
       adsurl = {https://ui.adsabs.harvard.edu/abs/2020ApJ...897...62F}
}

@ARTICLE{Kuriki2018ApJ...864..161K,
       author = {{Kuriki}, M. and {Sano}, H. and {Kuno}, N. and {Seta}, M. and {Yamane}, Y. and {Inaba}, T. and {Nagaya}, T. and {Yoshiike}, S. and {Okawa}, K. and {Tsutsumi}, D. and {Hattori}, Y. and {Kohno}, M. and {Fujita}, S. and {Nishimura}, A. and {Ohama}, A. and {Matsuo}, M. and {Tsuda}, Y. and {Torii}, K. and {Minamidani}, T. and {Umemoto}, T. and {Rowell}, G. and {Bamba}, A. and {Tachihara}, K. and {Fukui}, Y.},
        title = "{Discovery of Molecular and Atomic Clouds Associated with the Gamma-Ray Supernova Remnant Kesteven 79}",
      journal = {\apj},
         year = 2018,
        month = sep,
       volume = {864},
       number = {2},
          eid = {161},
        pages = {161},
          doi = {10.3847/1538-4357/aad7be},
archivePrefix = {arXiv},
       eprint = {1711.08165},
 primaryClass = {astro-ph.GA},
       adsurl = {https://ui.adsabs.harvard.edu/abs/2018ApJ...864..161K}
}

@ARTICLE{Landecker1989MNRAS.237..277L,
       author = {{Landecker}, T.~L. and {Pineault}, S. and {Routledge}, D. and {Vaneldik}, J.~F.},
        title = "{The interaction of the supernova remnant VRO 42.05.01 with its HI environment.}",
      journal = {\mnras},
         year = 1989,
        month = mar,
       volume = {237},
        pages = {277-297},
          doi = {10.1093/mnras/237.1.277},
       adsurl = {https://ui.adsabs.harvard.edu/abs/1989MNRAS.237..277L}
}

@ARTICLE{Telezhinsky2012APh....35..300T,
       author = {{Telezhinsky}, I. and {Dwarkadas}, V.~V. and {Pohl}, M.},
        title = "{Particle spectra from acceleration at forward and reverse shocks of young Type Ia Supernova Remnants}",
      journal = {Astroparticle Physics},
         year = 2012,
        month = jan,
       volume = {35},
       number = {6},
        pages = {300-311},
          doi = {10.1016/j.astropartphys.2011.10.001},
archivePrefix = {arXiv},
       eprint = {1110.0361},
 primaryClass = {astro-ph.HE},
       adsurl = {https://ui.adsabs.harvard.edu/abs/2012APh....35..300T}
}

@INPROCEEDINGS{Gouda2012ASPC..458..417G,
       author = {{Gouda}, N.},
        title = "{Infrared Space Astrometry Missions {\ensuremath{\sim}} JASMINE Missions {\ensuremath{\sim}}}",
    booktitle = {Galactic Archaeology: Near-Field Cosmology and the Formation of the Milky Way},
         year = 2012,
       editor = {{Aoki}, W. and {Ishigaki}, M. and {Suda}, T. and {Tsujimoto}, T. and {Arimoto}, N.},
       series = {Astronomical Society of the Pacific Conference Series},
       volume = {458},
        month = aug,
        pages = {417},
       adsurl = {https://ui.adsabs.harvard.edu/abs/2012ASPC..458..417G}
}

@ARTICLE{Aharonian2008A&A...481..401A,
       author = {{Aharonian}, F. and {Akhperjanian}, A.~G. and {Bazer-Bachi}, A.~R. and {Behera}, B. and {Beilicke}, M. and {Benbow}, W. and {Berge}, D. and {Bernl{\"o}hr}, K. and {Boisson}, C. and {Bolz}, O. and {Borrel}, V. and {Braun}, I. and {Brion}, E. and {Brown}, A.~M. and {B{\"u}hler}, R. and {Bulik}, T. and {B{\"u}sching}, I. and {Boutelier}, T. and {Carrigan}, S. and {Chadwick}, P.~M. and {Chounet}, L. -M. and {Clapson}, A.~C. and {Coignet}, G. and {Cornils}, R. and {Costamante}, L. and {Degrange}, B. and {Dickinson}, H.~J. and {Djannati-Ata{\"\i}}, A. and {Domainko}, W. and {O'C. Drury}, L. and {Dubus}, G. and {Dyks}, J. and {Egberts}, K. and {Emmanoulopoulos}, D. and {Espigat}, P. and {Farnier}, C. and {Feinstein}, F. and {Fiasson}, A. and {F{\"o}rster}, A. and {Fontaine}, G. and {Fukui}, Y. and {Funk}, Seb. and {Funk}, S. and {F{\"u}{\ss}ling}, M. and {Gallant}, Y.~A. and {Giebels}, B. and {Glicenstein}, J.~F. and {Gl{\"u}ck}, B. and {Goret}, P. and {Hadjichristidis}, C. and {Hauser}, D. and {Hauser}, M. and {Heinzelmann}, G. and {Henri}, G. and {Hermann}, G. and {Hinton}, J.~A. and {Hoffmann}, A. and {Hofmann}, W. and {Holleran}, M. and {Hoppe}, S. and {Horns}, D. and {Jacholkowska}, A. and {de Jager}, O.~C. and {Kendziorra}, E. and {Kerschhaggl}, M. and {Kh{\'e}lifi}, B. and {Komin}, Nu. and {Kosack}, K. and {Lamanna}, G. and {Latham}, I.~J. and {Le Gallou}, R. and {Lemi{\`e}re}, A. and {Lemoine-Goumard}, M. and {Lenain}, J. -P. and {Lohse}, T. and {Martin}, J.~M. and {Martineau-Huynh}, O. and {Marcowith}, A. and {Masterson}, C. and {Maurin}, G. and {McComb}, T.~J.~L. and {Moderski}, R. and {Moriguchi}, Y. and {Moulin}, E. and {de Naurois}, M. and {Nedbal}, D. and {Nolan}, S.~J. and {Olive}, J. -P. and {Orford}, K.~J. and {Osborne}, J.~L. and {Ostrowski}, M. and {Panter}, M. and {Pedaletti}, G. and {Pelletier}, G. and {Petrucci}, P. -O. and {Pita}, S. and {P{\"u}hlhofer}, G. and {Punch}, M. and {Ranchon}, S. and {Raubenheimer}, B.~C. and {Raue}, M. and {Rayner}, S.~M. and {Reimer}, O. and {Renaud}, M. and {Ripken}, J. and {Rob}, L. and {Rolland}, L. and {Rosier-Lees}, S. and {Rowell}, G. and {Rudak}, B. and {Ruppel}, J. and {Sahakian}, V. and {Santangelo}, A. and {Saug{\'e}}, L. and {Schlenker}, S. and {Schlickeiser}, R. and {Schr{\"o}der}, R. and {Schwanke}, U. and {Schwarzburg}, S. and {Schwemmer}, S. and {Shalchi}, A. and {Sol}, H. and {Spangler}, D. and {Stawarz}, {\L}. and {Steenkamp}, R. and {Stegmann}, C. and {Superina}, G. and {Takeuchi}, T. and {Tam}, P.~H. and {Tavernet}, J. -P. and {Terrier}, R. and {van Eldik}, C. and {Vasileiadis}, G. and {Venter}, C. and {Vialle}, J.~P. and {Vincent}, P. and {Vivier}, M. and {V{\"o}lk}, H.~J. and {Volpe}, F. and {Wagner}, S.~J. and {Ward}, M.},
        title = "{Discovery of very high energy gamma-ray emission coincident with molecular clouds in the W 28 (G6.4-0.1) field}",
      journal = {\aap},
         year = 2008,
        month = apr,
       volume = {481},
       number = {2},
        pages = {401-410},
          doi = {10.1051/0004-6361:20077765},
archivePrefix = {arXiv},
       eprint = {0801.3555},
 primaryClass = {astro-ph},
       adsurl = {https://ui.adsabs.harvard.edu/abs/2008A&A...481..401A}
}

@ARTICLE{Abeysekara2017Sci...358..911A,
       author = {{Abeysekara}, A.~U. and {Albert}, A. and {Alfaro}, R. and {Alvarez}, C. and {{\'A}lvarez}, J.~D. and {Arceo}, R. and {Arteaga-Vel{\'a}zquez}, J.~C. and {Avila Rojas}, D. and {Ayala Solares}, H.~A. and {Barber}, A.~S. and {Bautista-Elivar}, N. and {Becerril}, A. and {Belmont-Moreno}, E. and {BenZvi}, S.~Y. and {Berley}, D. and {Bernal}, A. and {Braun}, J. and {Brisbois}, C. and {Caballero-Mora}, K.~S. and {Capistr{\'a}n}, T. and {Carrami{\~n}ana}, A. and {Casanova}, S. and {Castillo}, M. and {Cotti}, U. and {Cotzomi}, J. and {Couti{\~n}o de Le{\'o}n}, S. and {De Le{\'o}n}, C. and {De la Fuente}, E. and {Dingus}, B.~L. and {DuVernois}, M.~A. and {D{\'\i}az-V{\'e}lez}, J.~C. and {Ellsworth}, R.~W. and {Engel}, K. and {Enr{\'\i}quez-Rivera}, O. and {Fiorino}, D.~W. and {Fraija}, N. and {Garc{\'\i}a-Gonz{\'a}lez}, J.~A. and {Garfias}, F. and {Gerhardt}, M. and {Gonz{\'a}lez Mu{\~n}oz}, A. and {Gonz{\'a}lez}, M.~M. and {Goodman}, J.~A. and {Hampel-Arias}, Z. and {Harding}, J.~P. and {Hern{\'a}ndez}, S. and {Hern{\'a}ndez-Almada}, A. and {Hinton}, J. and {Hona}, B. and {Hui}, C.~M. and {H{\"u}ntemeyer}, P. and {Iriarte}, A. and {Jardin-Blicq}, A. and {Joshi}, V. and {Kaufmann}, S. and {Kieda}, D. and {Lara}, A. and {Lauer}, R.~J. and {Lee}, W.~H. and {Lennarz}, D. and {Vargas}, H. Le{\'o}n and {Linnemann}, J.~T. and {Longinotti}, A.~L. and {Luis Raya}, G. and {Luna-Garc{\'\i}a}, R. and {L{\'o}pez-Coto}, R. and {Malone}, K. and {Marinelli}, S.~S. and {Martinez}, O. and {Martinez-Castellanos}, I. and {Mart{\'\i}nez-Castro}, J. and {Mart{\'\i}nez-Huerta}, H. and {Matthews}, J.~A. and {Miranda-Romagnoli}, P. and {Moreno}, E. and {Mostaf{\'a}}, M. and {Nellen}, L. and {Newbold}, M. and {Nisa}, M.~U. and {Noriega-Papaqui}, R. and {Pelayo}, R. and {Pretz}, J. and {P{\'e}rez-P{\'e}rez}, E.~G. and {Ren}, Z. and {Rho}, C.~D. and {Rivi{\`e}re}, C. and {Rosa-Gonz{\'a}lez}, D. and {Rosenberg}, M. and {Ruiz-Velasco}, E. and {Salazar}, H. and {Salesa Greus}, F. and {Sandoval}, A. and {Schneider}, M. and {Schoorlemmer}, H. and {Sinnis}, G. and {Smith}, A.~J. and {Springer}, R.~W. and {Surajbali}, P. and {Taboada}, I. and {Tibolla}, O. and {Tollefson}, K. and {Torres}, I. and {Ukwatta}, T.~N. and {Vianello}, G. and {Weisgarber}, T. and {Westerhoff}, S. and {Wisher}, I.~G. and {Wood}, J. and {Yapici}, T. and {Yodh}, G. and {Younk}, P.~W. and {Zepeda}, A. and {Zhou}, H. and {Guo}, F. and {Hahn}, J. and {Li}, H. and {Zhang}, H.},
        title = "{Extended gamma-ray sources around pulsars constrain the origin of the positron flux at Earth}",
      journal = {Science},
         year = 2017,
        month = nov,
       volume = {358},
       number = {6365},
        pages = {911-914},
          doi = {10.1126/science.aan4880},
archivePrefix = {arXiv},
       eprint = {1711.06223},
 primaryClass = {astro-ph.HE},
       adsurl = {https://ui.adsabs.harvard.edu/abs/2017Sci...358..911A}
}

@ARTICLE{Aharonian2008A&A...477..353A,
       author = {{Aharonian}, F. and {Akhperjanian}, A.~G. and {Barres de Almeida}, U. and {Bazer-Bachi}, A.~R. and {Behera}, B. and {Beilicke}, M. and {Benbow}, W. and {Bernl{\"o}hr}, K. and {Boisson}, C. and {Bolz}, O. and {Borrel}, V. and {Braun}, I. and {Brion}, E. and {Brown}, A.~M. and {B{\"u}hler}, R. and {Bulik}, T. and {B{\"u}sching}, I. and {Boutelier}, T. and {Carrigan}, S. and {Chadwick}, P.~M. and {Chounet}, L. -M. and {Clapson}, A.~C. and {Coignet}, G. and {Cornils}, R. and {Costamante}, L. and {Dalton}, M. and {Degrange}, B. and {Dickinson}, H.~J. and {Djannati-Ata{\"\i}}, A. and {Domainko}, W. and {Drury}, L. O'c. and {Dubois}, F. and {Dubus}, G. and {Dyks}, J. and {Egberts}, K. and {Emmanoulopoulos}, D. and {Espigat}, P. and {Farnier}, C. and {Feinstein}, F. and {Fiasson}, A. and {F{\"o}rster}, A. and {Fontaine}, G. and {Funk}, Seb. and {F{\"u}{\ss}ling}, M. and {Gallant}, Y.~A. and {Giebels}, B. and {Glicenstein}, J.~F. and {Gl{\"u}ck}, B. and {Goret}, P. and {Hadjichristidis}, C. and {Hauser}, D. and {Hauser}, M. and {Heinzelmann}, G. and {Henri}, G. and {Hermann}, G. and {Hinton}, J.~A. and {Hoffmann}, A. and {Hofmann}, W. and {Holleran}, M. and {Hoppe}, S. and {Horns}, D. and {Jacholkowska}, A. and {de Jager}, O.~C. and {Jung}, I. and {Katarzy{\'n}ski}, K. and {Kendziorra}, E. and {Kerschhaggl}, M. and {Kh{\'e}lifi}, B. and {Keogh}, D. and {Komin}, Nu. and {Kosack}, K. and {Lamanna}, G. and {Latham}, I.~J. and {Lemi{\`e}re}, A. and {Lemoine-Goumard}, M. and {Lenain}, J. -P. and {Lohse}, T. and {Martin}, J.~M. and {Martineau-Huynh}, O. and {Marcowith}, A. and {Masterson}, C. and {Maurin}, D. and {Maurin}, G. and {McComb}, T.~J.~L. and {Moderski}, R. and {Moulin}, E. and {de Naurois}, M. and {Nedbal}, D. and {Nolan}, S.~J. and {Ohm}, S. and {Olive}, J. -P. and {de O{\~n}a Wilhelmi}, E. and {Orford}, K.~J. and {Osborne}, J.~L. and {Ostrowski}, M. and {Panter}, M. and {Pedaletti}, G. and {Pelletier}, G. and {Petrucci}, P. -O. and {Pita}, S. and {P{\"u}hlhofer}, G. and {Punch}, M. and {Ranchon}, S. and {Raubenheimer}, B.~C. and {Raue}, M. and {Rayner}, S.~M. and {Renaud}, M. and {Ripken}, J. and {Rob}, L. and {Rolland}, L. and {Rosier-Lees}, S. and {Rowell}, G. and {Rudak}, B. and {Ruppel}, J. and {Sahakian}, V. and {Santangelo}, A. and {Schlickeiser}, R. and {Sch{\"o}ck}, F. and {Schr{\"o}der}, R. and {Schwanke}, U. and {Schwarzburg}, S. and {Schwemmer}, S. and {Shalchi}, A. and {Sol}, H. and {Spangler}, D. and {Stawarz}, {\L}. and {Steenkamp}, R. and {Stegmann}, C. and {Superina}, G. and {Tam}, P.~H. and {Tavernet}, J. -P. and {Terrier}, R. and {van Eldik}, C. and {Vasileiadis}, G. and {Venter}, C. and {Vialle}, J.~P. and {Vincent}, P. and {Vivier}, M. and {V{\"o}lk}, H.~J. and {Volpe}, F. and {Wagner}, S.~J. and {Ward}, M. and {Zdziarski}, A.~A. and {Zech}, A.},
        title = "{HESS very-high-energy gamma-ray sources without identified counterparts}",
      journal = {\aap},
         year = 2008,
        month = jan,
       volume = {477},
       number = {1},
        pages = {353-363},
          doi = {10.1051/0004-6361:20078516},
archivePrefix = {arXiv},
       eprint = {0712.1173},
 primaryClass = {astro-ph},
       adsurl = {https://ui.adsabs.harvard.edu/abs/2008A&A...477..353A}
}

@ARTICLE{Suzuki2022ApJ...924...45S,
       author = {{Suzuki}, Hiromasa and {Bamba}, Aya and {Yamazaki}, Ryo and {Ohira}, Yutaka},
        title = "{Observational Constraints on the Maximum Energies of Accelerated Particles in Supernova Remnants: Low Maximum Energies and a Large Variety}",
      journal = {\apj},
         year = 2022,
        month = jan,
       volume = {924},
       number = {2},
          eid = {45},
        pages = {45},
          doi = {10.3847/1538-4357/ac33b5},
archivePrefix = {arXiv},
       eprint = {2110.13304},
 primaryClass = {astro-ph.HE},
       adsurl = {https://ui.adsabs.harvard.edu/abs/2022ApJ...924...45S}
}

@ARTICLE{Aharonian2006A&A...449..223A,
       author = {{Aharonian}, F. and {Akhperjanian}, A.~G. and {Bazer-Bachi}, A.~R. and {Beilicke}, M. and {Benbow}, W. and {Berge}, D. and {Bernl{\"o}hr}, K. and {Boisson}, C. and {Bolz}, O. and {Borrel}, V. and {Braun}, I. and {Breitling}, F. and {Brown}, A.~M. and {Chadwick}, P.~M. and {Chounet}, L. -M. and {Cornils}, R. and {Costamante}, L. and {Degrange}, B. and {Dickinson}, H.~J. and {Djannati-Ata{\"\i}}, A. and {O'C. Drury}, L. and {Dubus}, G. and {Emmanoulopoulos}, D. and {Espigat}, P. and {Feinstein}, F. and {Fontaine}, G. and {Fuchs}, Y. and {Funk}, S. and {Gallant}, Y.~A. and {Giebels}, B. and {Glicenstein}, J.~F. and {Goret}, P. and {Hadjichristidis}, C. and {Hauser}, D. and {Hauser}, M. and {Heinzelmann}, G. and {Henri}, G. and {Hermann}, G. and {Hinton}, J.~A. and {Hofmann}, W. and {Holleran}, M. and {Horns}, D. and {Jacholkowska}, A. and {de Jager}, O.~C. and {Kh{\'e}lifi}, B. and {Klages}, S. and {Komin}, Nu. and {Konopelko}, A. and {Latham}, I.~J. and {Le Gallou}, R. and {Lemi{\`e}re}, A. and {Lemoine-Goumard}, M. and {Lohse}, T. and {Martin}, J.~M. and {Martineau-Huynh}, O. and {Marcowith}, A. and {Masterson}, C. and {McComb}, T.~J.~L. and {de Naurois}, M. and {Nedbal}, D. and {Nolan}, S.~J. and {Noutsos}, A. and {Orford}, K.~J. and {Osborne}, J.~L. and {Ouchrif}, M. and {Panter}, M. and {Pelletier}, G. and {Pita}, S. and {P{\"u}hlhofer}, G. and {Punch}, M. and {Raubenheimer}, B.~C. and {Raue}, M. and {Rayner}, S.~M. and {Reimer}, A. and {Reimer}, O. and {Ripken}, J. and {Rob}, L. and {Rolland}, L. and {Rowell}, G. and {Sahakian}, V. and {Saug{\'e}}, L. and {Schlenker}, S. and {Schlickeiser}, R. and {Schuster}, C. and {Schwanke}, U. and {Siewert}, M. and {Sol}, H. and {Spangler}, D. and {Steenkamp}, R. and {Stegmann}, C. and {Superina}, G. and {Tavernet}, J. -P. and {Terrier}, R. and {Th{\'e}oret}, C.~G. and {Tluczykont}, M. and {van Eldik}, C. and {Vasileiadis}, G. and {Venter}, C. and {Vincent}, P. and {V{\"o}lk}, H.~J. and {Wagner}, S.~J.},
        title = "{A detailed spectral and morphological study of the gamma-ray supernova remnant <ASTROBJ>RX J1713.7-3946</ASTROBJ> with HESS}",
      journal = {\aap},
         year = 2006,
        month = apr,
       volume = {449},
       number = {1},
        pages = {223-242},
          doi = {10.1051/0004-6361:20054279},
archivePrefix = {arXiv},
       eprint = {astro-ph/0511678},
 primaryClass = {astro-ph},
       adsurl = {https://ui.adsabs.harvard.edu/abs/2006A&A...449..223A}
}

@ARTICLE{Aharonian1996A&A...309..917A,
       author = {{Aharonian}, F.~A. and {Atoyan}, A.~M.},
        title = "{On the emissivity of {\ensuremath{\pi}}\^0\^-decay gamma radiation in the vicinity of accelerators of galactic cosmic rays.}",
      journal = {\aap},
         year = 1996,
        month = may,
       volume = {309},
        pages = {917-928},
       adsurl = {https://ui.adsabs.harvard.edu/abs/1996A&A...309..917A}
}

@ARTICLE{Ohira2010A&A...513A..17O,
       author = {{Ohira}, Y. and {Murase}, K. and {Yamazaki}, R.},
        title = "{Escape-limited model of cosmic-ray acceleration revisited}",
      journal = {\aap},
         year = 2010,
        month = apr,
       volume = {513},
          eid = {A17},
        pages = {A17},
          doi = {10.1051/0004-6361/200913495},
archivePrefix = {arXiv},
       eprint = {0910.3449},
 primaryClass = {astro-ph.HE},
       adsurl = {https://ui.adsabs.harvard.edu/abs/2010A&A...513A..17O}
}

@ARTICLE{Sato1978AJ.....83.1607S,
       author = {{Sato}, F. and {Fukui}, Y.},
        title = "{Atomic hydrogen in the giant molecular cloud near M17.}",
      journal = {\aj},
         year = 1978,
        month = dec,
       volume = {83},
        pages = {1607-1611},
          doi = {10.1086/112370},
       adsurl = {https://ui.adsabs.harvard.edu/abs/1978AJ.....83.1607S}
}

@ARTICLE{Dickey1990ARA&A..28..215D,
       author = {{Dickey}, John M. and {Lockman}, Felix J.},
        title = "{H I in the galaxy.}",
      journal = {\araa},
         year = 1990,
        month = jan,
       volume = {28},
        pages = {215-261},
          doi = {10.1146/annurev.aa.28.090190.001243},
       adsurl = {https://ui.adsabs.harvard.edu/abs/1990ARA&A..28..215D}
}

@ARTICLE{Okamoto2017ApJ...838..132O,
       author = {{Okamoto}, Ryuji and {Yamamoto}, Hiroaki and {Tachihara}, Kengo and {Hayakawa}, Takahiro and {Hayashi}, Katsuhiro and {Fukui}, Yasuo},
        title = "{H I, CO, and Dust in the Perseus Cloud}",
      journal = {\apj},
         year = 2017,
        month = apr,
       volume = {838},
       number = {2},
          eid = {132},
        pages = {132},
          doi = {10.3847/1538-4357/aa6747},
archivePrefix = {arXiv},
       eprint = {1612.07696},
 primaryClass = {astro-ph.GA},
       adsurl = {https://ui.adsabs.harvard.edu/abs/2017ApJ...838..132O}
}

@ARTICLE{Albert2006ApJ...643L..53A,
       author = {{Albert}, J. and {Aliu}, E. and {Anderhub}, H. and {Antoranz}, P. and {Armada}, A. and {Asensio}, M. and {Baixeras}, C. and {Barrio}, J.~A. and {Bartelt}, M. and {Bartko}, H. and {Bastieri}, D. and {Bavikadi}, S.~R. and {Bednarek}, W. and {Berger}, K. and {Bigongiari}, C. and {Biland}, A. and {Bisesi}, E. and {Bock}, R.~K. and {Bordas}, P. and {Bosch-Ramon}, V. and {Bretz}, T. and {Britvitch}, I. and {Camara}, M. and {Carmona}, E. and {Chilingarian}, A. and {Ciprini}, S. and {Coarasa}, J.~A. and {Commichau}, S. and {Contreras}, J.~L. and {Cortina}, J. and {Curtef}, V. and {Dame}, T.~M. and {Danielyan}, V. and {Dazzi}, F. and {De Angelis}, A. and {de los Reyes}, R. and {De Lotto}, B. and {Domingo-Santamar{\'\i}}, E. and {Dorner}, D. and {Doro}, M. and {Errando}, M. and {Fagiolini}, M. and {Ferenc}, D. and {Fern{\'a}ndez}, E. and {Firpo}, R. and {Flix}, J. and {Fonseca}, M.~V. and {Font}, L. and {Fuchs}, M. and {Galante}, N. and {Garczarczyk}, M. and {Gaug}, M. and {Giller}, M. and {Goebel}, F. and {Hakobyan}, D. and {Hayashida}, M. and {Hengstebeck}, T. and {H{\"o}hne}, D. and {Hose}, J. and {Hsu}, C.~C. and {Isar}, P.~G. and {Jacon}, P. and {Kalekin}, O. and {Kasyra}, R. and {Kranich}, D. and {Laatiaoui}, M. and {Laille}, A. and {Lenisa}, T. and {Liebing}, P. and {Lindfors}, E. and {Lombardi}, S. and {Longo}, F. and {L{\'o}pez}, J. and {L{\'o}pez}, M. and {Lorenz}, E. and {Lucarelli}, F. and {Majumdar}, P. and {Maneva}, G. and {Mannheim}, K. and {Mansutti}, O. and {Mariotti}, M. and {Mart{\'\i}nez}, M. and {Mase}, K. and {Mazin}, D. and {Merck}, C. and {Meucci}, M. and {Meyer}, M. and {Miranda}, J.~M. and {Mirzoyan}, R. and {Mizobuchi}, S. and {Moralejo}, A. and {Nilsson}, K. and {O{\~n}a-Wilhelmi}, E. and {Ordu{\~n}a}, R. and {Otte}, N. and {Oya}, I. and {Paneque}, D. and {Paoletti}, R. and {Paredes}, J.~M. and {Pasanen}, M. and {Pascoli}, D. and {Pauss}, F. and {Pavel}, N. and {Pegna}, R. and {Persic}, M. and {Peruzzo}, L. and {Piccioli}, A. and {Poller}, M. and {Prandini}, E. and {Raymers}, A. and {Rico}, J. and {Rhode}, W. and {Rib{\'o}}, M. and {Riegel}, B. and {Rissi}, M. and {Robert}, A. and {R{\"u}gamer}, S. and {Saggion}, A. and {S{\'a}nchez}, A. and {Sartori}, P. and {Scalzotto}, V. and {Scapin}, V. and {Schmitt}, R. and {Schweizer}, T. and {Shayduk}, M. and {Shinozaki}, K. and {Shore}, S.~N. and {Sidro}, N. and {Sillanp{\"a}{\"a}}, A. and {Sobczynska}, D. and {Stamerra}, A. and {Stark}, L.~S. and {Takalo}, L. and {Temnikov}, P. and {Tescaro}, D. and {Teshima}, M. and {Tonello}, N. and {Torres}, A. and {Torres}, D.~F. and {Turini}, N. and {Vankov}, H. and {Vitale}, V. and {Wagner}, R.~M. and {Wibig}, T. and {Wittek}, W. and {Zanin}, R. and {Zapatero}, J.},
        title = "{Observation of VHE Gamma Radiation from HESS J1834-087/W41 with the MAGIC Telescope}",
      journal = {\apjl},
         year = 2006,
        month = may,
       volume = {643},
       number = {1},
        pages = {L53-L56},
          doi = {10.1086/504917},
archivePrefix = {arXiv},
       eprint = {astro-ph/0604197},
 primaryClass = {astro-ph},
       adsurl = {https://ui.adsabs.harvard.edu/abs/2006ApJ...643L..53A}
}

@ARTICLE{Arikawa1999PASJ...51L...7A,
       author = {{Arikawa}, Yuji and {Tatematsu}, Ken'ichi and {Sekimoto}, Yutaro and {Takahashi}, Tadayuki},
        title = "{Shocked Molecular Gas Associated with the Supernova Remnant W28}",
      journal = {\pasj},
         year = 1999,
        month = aug,
       volume = {51},
        pages = {L7-L10},
          doi = {10.1093/pasj/51.4.L7},
       adsurl = {https://ui.adsabs.harvard.edu/abs/1999PASJ...51L...7A}
}

@ARTICLE{Broersen2014MNRAS.441.3040B,
       author = {{Broersen}, Sjors and {Chiotellis}, Alexandros and {Vink}, Jacco and {Bamba}, Aya},
        title = "{The many sides of RCW 86: a Type Ia supernova remnant evolving in its progenitor's wind bubble}",
      journal = {\mnras},
         year = 2014,
        month = jul,
       volume = {441},
       number = {4},
        pages = {3040-3054},
          doi = {10.1093/mnras/stu667},
archivePrefix = {arXiv},
       eprint = {1404.5434},
 primaryClass = {astro-ph.HE},
       adsurl = {https://ui.adsabs.harvard.edu/abs/2014MNRAS.441.3040B}
}

@ARTICLE{Weaver1977ApJ...218..377W,
       author = {{Weaver}, R. and {McCray}, R. and {Castor}, J. and {Shapiro}, P. and {Moore}, R.},
        title = "{Interstellar bubbles. II. Structure and evolution.}",
      journal = {\apj},
         year = 1977,
        month = dec,
       volume = {218},
        pages = {377-395},
          doi = {10.1086/155692},
       adsurl = {https://ui.adsabs.harvard.edu/abs/1977ApJ...218..377W}
}

@ARTICLE{Dwarkadas2005ApJ...630..892D,
       author = {{Dwarkadas}, Vikram V.},
        title = "{The Evolution of Supernovae in Circumstellar Wind-Blown Bubbles. I. Introduction and One-Dimensional Calculations}",
      journal = {\apj},
         year = 2005,
        month = sep,
       volume = {630},
       number = {2},
        pages = {892-910},
          doi = {10.1086/432109},
archivePrefix = {arXiv},
       eprint = {astro-ph/0410464},
 primaryClass = {astro-ph},
       adsurl = {https://ui.adsabs.harvard.edu/abs/2005ApJ...630..892D}
}

@ARTICLE{Koo1992ApJ...388...93K,
       author = {{Koo}, Bon-Chul and {McKee}, Christopher F.},
        title = "{Dynamics of Wind Bubbles and Superbubbles. I. Slow Winds and Fast Winds}",
      journal = {\apj},
         year = 1992,
        month = mar,
       volume = {388},
        pages = {93},
          doi = {10.1086/171132},
       adsurl = {https://ui.adsabs.harvard.edu/abs/1992ApJ...388...93K}
}

@ARTICLE{Koo1991ApJ...382..204K,
       author = {{Koo}, Bon-Chul and {Heiles}, Carl},
        title = "{A Survey of H i 21 Centimeter Emission Lines toward Supernova Remnants}",
      journal = {\apj},
         year = 1991,
        month = nov,
       volume = {382},
        pages = {204},
          doi = {10.1086/170709},
       adsurl = {https://ui.adsabs.harvard.edu/abs/1991ApJ...382..204K}
}

@ARTICLE{Koo1990ApJ...364..178K,
       author = {{Koo}, Bon-Chul and {Reach}, William T. and {Heiles}, Carl and {Fesen}, Robert A. and {Shull}, J. Michael},
        title = "{Detection of an Expanding H i Shell in the Old Supernova Remnant CTB 80}",
      journal = {\apj},
         year = 1990,
        month = nov,
       volume = {364},
        pages = {178},
          doi = {10.1086/169400},
       adsurl = {https://ui.adsabs.harvard.edu/abs/1990ApJ...364..178K}
}

@ARTICLE{Inoue2009ApJ...695..825I,
       author = {{Inoue}, Tsuyoshi and {Yamazaki}, Ryo and {Inutsuka}, Shu-ichiro},
        title = "{Turbulence and Magnetic Field Amplification in Supernova Remnants: Interactions Between a Strong Shock Wave and Multiphase Interstellar Medium}",
      journal = {\apj},
         year = 2009,
        month = apr,
       volume = {695},
       number = {2},
        pages = {825-833},
          doi = {10.1088/0004-637X/695/2/825},
archivePrefix = {arXiv},
       eprint = {0901.0486},
 primaryClass = {astro-ph.SR},
       adsurl = {https://ui.adsabs.harvard.edu/abs/2009ApJ...695..825I}
}

@ARTICLE{Inoue2012ApJ...744...71I,
       author = {{Inoue}, Tsuyoshi and {Yamazaki}, Ryo and {Inutsuka}, Shu-ichiro and {Fukui}, Yasuo},
        title = "{Toward Understanding the Cosmic-Ray Acceleration at Young Supernova Remnants Interacting with Interstellar Clouds: Possible Applications to RX J1713.7-3946}",
      journal = {\apj},
         year = 2012,
        month = jan,
       volume = {744},
       number = {1},
          eid = {71},
        pages = {71},
          doi = {10.1088/0004-637X/744/1/71},
archivePrefix = {arXiv},
       eprint = {1106.3080},
 primaryClass = {astro-ph.HE},
       adsurl = {https://ui.adsabs.harvard.edu/abs/2012ApJ...744...71I}
}

@ARTICLE{Hess1912,
       author = {{Hess}, V. },
        title = "{Über Beobachtungen der durchdringenden Strahlung bei sieben Freiballonfahrten}",
      journal = {Phys.Z.},
         year = 1912,
       volume = {13},
        pages = {1084},
}

@ARTICLE{Messineo2010ApJ...708.1241M,
       author = {{Messineo}, Maria and {Figer}, Donald F. and {Davies}, Ben and {Kudritzki}, R.~P. and {Rich}, R. Michael and {MacKenty}, John and {Trombley}, Christine},
        title = "{Hubble Space Telescope/Near-Infrared Camera and Multi-Object Spectrometer Observations of the GLIMPSE9 Stellar Cluster}",
      journal = {\apj},
         year = 2010,
        month = jan,
       volume = {708},
       number = {2},
        pages = {1241-1253},
          doi = {10.1088/0004-637X/708/2/1241},
archivePrefix = {arXiv},
       eprint = {0912.2061},
 primaryClass = {astro-ph.GA},
       adsurl = {https://ui.adsabs.harvard.edu/abs/2010ApJ...708.1241M}
}

@ARTICLE{Thompson2006A&A...453.1003T,
       author = {{Thompson}, M.~A. and {Hatchell}, J. and {Walsh}, A.~J. and {MacDonald}, G.~H. and {Millar}, T.~J.},
        title = "{A SCUBA imaging survey of ultracompact HII regions. The environments of massive star formation}",
      journal = {\aap},
         year = 2006,
        month = jul,
       volume = {453},
       number = {3},
        pages = {1003-1026},
          doi = {10.1051/0004-6361:20054383},
archivePrefix = {arXiv},
       eprint = {astro-ph/0604208},
 primaryClass = {astro-ph},
       adsurl = {https://ui.adsabs.harvard.edu/abs/2006A&A...453.1003T}
}

@ARTICLE{Chibueze2025MNRAS.539..145C,
       author = {{Chibueze}, James O. and {Ugwu}, Chukwuebuka J. and {Hirota}, Tomoya and {Kim}, Kee-Tae and {Liu}, Tie and {Vorster}, Jakobus M. and {Kang}, Ji-hyun and {Kim}, Jungha and {Burns}, Ross A. and {Sobolev}, Andrey M. and {Hwang}, Jihye and {Lee}, Chang Won and {Kim}, Mi Kyoung and {Sugiyama}, Koichiro},
        title = "{Spiral arm, rotating structure, and outflow cavity in massive star-forming region G23.43-0.18}",
      journal = {\mnras},
         year = 2025,
        month = may,
       volume = {539},
       number = {1},
        pages = {145-159},
          doi = {10.1093/mnras/stae2773},
       adsurl = {https://ui.adsabs.harvard.edu/abs/2025MNRAS.539..145C}
}

@ARTICLE{Fujisawa2014PASJ...66...31F,
       author = {{Fujisawa}, Kenta and {Sugiyama}, Koichiro and {Motogi}, Kazuhito and {Hachisuka}, Kazuya and {Yonekura}, Yoshinori and {Sawada-Satoh}, Satoko and {Matsumoto}, Naoko and {Sorai}, Kazuo and {Momose}, Munetake and {Saito}, Yu and {Takaba}, Hiroshi and {Ogawa}, Hideo and {Kimura}, Kimihiro and {Niinuma}, Kotaro and {Hirano}, Daiki and {Omodaka}, Toshihiro and {Kobayashi}, Hideyuki and {Kawaguchi}, Noriyuki and {Shibata}, Katsunori M. and {Honma}, Mareki and {Hirota}, Tomoya and {Murata}, Yasuhiro and {Doi}, Akihiro and {Mochizuki}, Nanako and {Shen}, Zhiqiang and {Chen}, Xi and {Xia}, Bo and {Li}, Bin and {Kim}, Kee-Tae},
        title = "{Observations of 6.7 GHz methanol masers with East-Asian VLBI Network. I. VLBI images of the first epoch of observations}",
      journal = {\pasj},
         year = 2014,
        month = apr,
       volume = {66},
       number = {2},
          eid = {31},
        pages = {31},
          doi = {10.1093/pasj/psu015},
archivePrefix = {arXiv},
       eprint = {1311.3431},
 primaryClass = {astro-ph.SR},
       adsurl = {https://ui.adsabs.harvard.edu/abs/2014PASJ...66...31F}
}

@ARTICLE{Brunthaler2009ApJ...693..424B,
       author = {{Brunthaler}, A. and {Reid}, M.~J. and {Menten}, K.~M. and {Zheng}, X.~W. and {Moscadelli}, L. and {Xu}, Y.},
        title = "{Trigonometric Parallaxes of Massive Star-Forming Regions. V. G23.01-0.41 and G23.44-0.18}",
      journal = {\apj},
         year = 2009,
        month = mar,
       volume = {693},
       number = {1},
        pages = {424-429},
          doi = {10.1088/0004-637X/693/1/424},
archivePrefix = {arXiv},
       eprint = {0811.0713},
 primaryClass = {astro-ph},
       adsurl = {https://ui.adsabs.harvard.edu/abs/2009ApJ...693..424B}
}

@ARTICLE{Su2014ApJ...796..122S,
       author = {{Su}, Yang and {Yang}, Ji and {Zhou}, Xin and {Zhou}, Ping and {Chen}, Yang},
        title = "{Interaction between Supernova Remnant G22.7-0.2 and the Ambient Molecular Clouds}",
      journal = {\apj},
         year = 2014,
        month = dec,
       volume = {796},
       number = {2},
          eid = {122},
        pages = {122},
          doi = {10.1088/0004-637X/796/2/122},
archivePrefix = {arXiv},
       eprint = {1411.0757},
 primaryClass = {astro-ph.SR},
       adsurl = {https://ui.adsabs.harvard.edu/abs/2014ApJ...796..122S}
}

@ARTICLE{Su2015ApJ...811..134S,
       author = {{Su}, Yang and {Zhang}, Shaobo and {Shao}, Xiangjun and {Yang}, Ji},
        title = "{The Dense Filamentary Giant Molecular Cloud G23.0-0.4: Birthplace of Ongoing Massive Star Formation}",
      journal = {\apj},
         year = 2015,
        month = oct,
       volume = {811},
       number = {2},
          eid = {134},
        pages = {134},
          doi = {10.1088/0004-637X/811/2/134},
archivePrefix = {arXiv},
       eprint = {1508.07898},
 primaryClass = {astro-ph.GA},
       adsurl = {https://ui.adsabs.harvard.edu/abs/2015ApJ...811..134S}
}

@ARTICLE{Leahy2008AJ....135..167L,
       author = {{Leahy}, D.~A. and {Tian}, W.~W.},
        title = "{The Distances of SNR W41 and Overlapping H II Regions}",
      journal = {\aj},
         year = 2008,
        month = jan,
       volume = {135},
       number = {1},
        pages = {167-172},
          doi = {10.1088/0004-6256/135/1/167},
archivePrefix = {arXiv},
       eprint = {0708.3377},
 primaryClass = {astro-ph},
       adsurl = {https://ui.adsabs.harvard.edu/abs/2008AJ....135..167L}
}

@ARTICLE{Tam2020ApJ...899...75T,
       author = {{Tam}, Pak-Hin Thomas and {Lee}, K.~K. and {Cui}, Yudong and {Hu}, C.~P. and {Kong}, A.~K.~H. and {Li}, K.~L. and {Tudor}, Vlad and {He}, Xinbo and {Pal}, Partha S.},
        title = "{A Multiwavelength Study of the {\ensuremath{\gamma}}-Ray Binary Candidate HESS J1832-093}",
      journal = {\apj},
         year = 2020,
        month = aug,
       volume = {899},
       number = {1},
          eid = {75},
        pages = {75},
          doi = {10.3847/1538-4357/ab9e76},
archivePrefix = {arXiv},
       eprint = {2001.07138},
 primaryClass = {astro-ph.HE},
       adsurl = {https://ui.adsabs.harvard.edu/abs/2020ApJ...899...75T}
}

@ARTICLE{Castro2013ApJ...774...36C,
       author = {{Castro}, Daniel and {Slane}, Patrick and {Carlton}, Ashley and {Figueroa-Feliciano}, Enectali},
        title = "{Fermi-LAT Observations of Supernova Remnants Interacting with Molecular Clouds: W41, MSH 17-39, and G337.7-0.1}",
      journal = {\apj},
         year = 2013,
        month = sep,
       volume = {774},
       number = {1},
          eid = {36},
        pages = {36},
          doi = {10.1088/0004-637X/774/1/36},
archivePrefix = {arXiv},
       eprint = {1305.3623},
 primaryClass = {astro-ph.HE},
       adsurl = {https://ui.adsabs.harvard.edu/abs/2013ApJ...774...36C}
}

@ARTICLE{HESS2015MNRAS.446.1163H,
       author = {{H.~E.~S.~S. Collaboration} and {Abramowski}, A. and {Acero}, F. and {Aharonian}, F. and {Ait Benkhali}, F. and {Akhperjanian}, A.~G. and {Ang{\"u}ner}, E. and {Anton}, G. and {Balenderan}, S. and {Balzer}, A. and {Barnacka}, A. and {Becherini}, Y. and {Becker Tjus}, J. and {Bernl{\"o}hr}, K. and {Birsin}, E. and {Bissaldi}, E. and {Biteau}, J. and {B{\"o}ttcher}, M. and {Boisson}, C. and {Bolmont}, J. and {Bordas}, P. and {Brucker}, J. and {Brun}, F. and {Brun}, P. and {Bulik}, T. and {Carrigan}, S. and {Casanova}, S. and {Cerruti}, M. and {Chadwick}, P.~M. and {Chalme-Calvet}, R. and {Chaves}, R.~C.~G. and {Cheesebrough}, A. and {Chr{\'e}tien}, M. and {Clapson}, A. -C. and {Colafrancesco}, S. and {Cologna}, G. and {Conrad}, J. and {Couturier}, C. and {Cui}, Y. and {Dalton}, M. and {Daniel}, M.~K. and {Davids}, I.~D. and {Degrange}, B. and {Deil}, C. and {deWilt}, P. and {Dickinson}, H.~J. and {Djannati-Ata{\"\i}}, A. and {Domainko}, W. and {Drury}, L. O'C. and {Dubus}, G. and {Dutson}, K. and {Dyks}, J. and {Dyrda}, M. and {Edwards}, T. and {Egberts}, K. and {Eger}, P. and {Espigat}, P. and {Farnier}, C. and {Fegan}, S. and {Feinstein}, F. and {Fernandes}, M.~V. and {Fernandez}, D. and {Fiasson}, A. and {Fontaine}, G. and {F{\"o}rster}, A. and {F{\"u}{\ss}ling}, M. and {Gajdus}, M. and {Gallant}, Y.~A. and {Garrigoux}, T. and {Giavitto}, G. and {Giebels}, B. and {Glicenstein}, J.~F. and {Grondin}, M. -H. and {Grudzi{\'n}ska}, M. and {H{\"a}ffner}, S. and {Hahn}, J. and {Harris}, J. and {Heinzelmann}, G. and {Henri}, G. and {Hermann}, G. and {Hervet}, O. and {Hillert}, A. and {Hinton}, J.~A. and {Hofmann}, W. and {Hofverberg}, P. and {Holler}, M. and {Horns}, D. and {Jacholkowska}, A. and {Jahn}, C. and {Jamrozy}, M. and {Janiak}, M. and {Jankowsky}, F. and {Jung}, I. and {Kastendieck}, M.~A. and {Katarzy{\'n}ski}, K. and {Katz}, U. and {Kaufmann}, S. and {Kh{\'e}lifi}, B. and {Kieffer}, M. and {Klepser}, S. and {Klochkov}, D. and {Klu{\'z}niak}, W. and {Kneiske}, T. and {Kolitzus}, D. and {Komin}, Nu. and {Kosack}, K. and {Krakau}, S. and {Krayzel}, F. and {Kr{\"u}ger}, P.~P. and {Laffon}, H. and {Lamanna}, G. and {Lefaucheur}, J. and {Lemi{\`e}re}, A. and {Lemoine-Goumard}, M. and {Lenain}, J. -P. and {Lennarz}, D. and {Lohse}, T. and {Lopatin}, A. and {Lu}, C. -C. and {Marandon}, V. and {Marcowith}, A. and {Marx}, R. and {Maurin}, G. and {Maxted}, N. and {Mayer}, M. and {McComb}, T.~J.~L. and {M{\'e}hault}, J. and {Meintjes}, P.~J. and {Menzler}, U. and {Meyer}, M. and {Moderski}, R. and {Mohamed}, M. and {Moulin}, E. and {Murach}, T. and {Naumann}, C.~L. and {de Naurois}, M. and {Niemiec}, J. and {Nolan}, S.~J. and {Oakes}, L. and {Ohm}, S. and {de O{\~n}a Wilhelmi}, E. and {Opitz}, B. and {Ostrowski}, M. and {Oya}, I. and {Panter}, M. and {Parsons}, R.~D. and {Paz Arribas}, M. and {Pekeur}, N.~W. and {Pelletier}, G. and {Perez}, J. and {Petrucci}, P. -O. and {Peyaud}, B. and {Pita}, S. and {Poon}, H. and {P{\"u}hlhofer}, G. and {Punch}, M. and {Quirrenbach}, A. and {Raab}, S. and {Raue}, M. and {Reimer}, A. and {Reimer}, O. and {Renaud}, M. and {de los Reyes}, R. and {Rieger}, F. and {Rob}, L. and {Romoli}, C. and {Rosier-Lees}, S. and {Rowell}, G. and {Rudak}, B. and {Rulten}, C.~B. and {Sahakian}, V. and {Sanchez}, D.~A. and {Santangelo}, A. and {Schlickeiser}, R. and {Sch{\"u}ssler}, F. and {Schulz}, A. and {Schwanke}, U. and {Schwarzburg}, S. and {Schwemmer}, S. and {Sol}, H. and {Spengler}, G. and {Spies}, F. and {Stawarz}, {\L}. and {Steenkamp}, R. and {Stegmann}, C. and {Stinzing}, F. and {Stycz}, K. and {Sushch}, I. and {Szostek}, A. and {Tavernet}, J. -P. and {Tavernier}, T. and {Taylor}, A.~M. and {Terrier}, R. and {Tluczykont}, M. and {Trichard}, C. and {Valerius}, K. and {van Eldik}, C. and {van Soelen}, B. and {Vasileiadis}, G. and {Venter}, C.},
        title = "{Discovery of the VHE gamma-ray source HESS J1832-093 in the vicinity of SNR G22.7-0.2}",
      journal = {\mnras},
         year = 2015,
        month = jan,
       volume = {446},
       number = {2},
        pages = {1163-1169},
          doi = {10.1093/mnras/stu2148},
archivePrefix = {arXiv},
       eprint = {1411.0572},
 primaryClass = {astro-ph.HE},
       adsurl = {https://ui.adsabs.harvard.edu/abs/2015MNRAS.446.1163H}
}

@ARTICLE{Stafford2019ApJ...884..113S,
       author = {{Stafford}, Jennifer N. and {Lopez}, Laura A. and {Auchettl}, Katie and {Holland-Ashford}, Tyler},
        title = "{The Age Evolution of the Radio Morphology of Supernova Remnants}",
      journal = {\apj},
         year = 2019,
        month = oct,
       volume = {884},
       number = {2},
          eid = {113},
        pages = {113},
          doi = {10.3847/1538-4357/ab3a33},
archivePrefix = {arXiv},
       eprint = {1808.08234},
 primaryClass = {astro-ph.HE},
       adsurl = {https://ui.adsabs.harvard.edu/abs/2019ApJ...884..113S}
}

@ARTICLE{Tsuji2021ApJ...907..117T,
       author = {{Tsuji}, Naomi and {Uchiyama}, Yasunobu and {Khangulyan}, Dmitry and {Aharonian}, Felix},
        title = "{Systematic Study of Acceleration Efficiency in Young Supernova Remnants with Nonthermal X-Ray Observations}",
      journal = {\apj},
         year = 2021,
        month = feb,
       volume = {907},
       number = {2},
          eid = {117},
        pages = {117},
          doi = {10.3847/1538-4357/abce65},
archivePrefix = {arXiv},
       eprint = {2012.01047},
 primaryClass = {astro-ph.HE},
       adsurl = {https://ui.adsabs.harvard.edu/abs/2021ApJ...907..117T}
}

@ARTICLE{Uchiyama2012ApJ...749L..35U,
       author = {{Uchiyama}, Yasunobu and {Funk}, Stefan and {Katagiri}, Hideaki and {Katsuta}, Junichiro and {Lemoine-Goumard}, Marianne and {Tajima}, Hiroyasu and {Tanaka}, Takaaki and {Torres}, Diego F.},
        title = "{Fermi Large Area Telescope Discovery of GeV Gamma-Ray Emission from the Vicinity of SNR W44}",
      journal = {\apjl},
         year = 2012,
        month = apr,
       volume = {749},
       number = {2},
          eid = {L35},
        pages = {L35},
          doi = {10.1088/2041-8205/749/2/L35},
archivePrefix = {arXiv},
       eprint = {1203.3234},
 primaryClass = {astro-ph.HE},
       adsurl = {https://ui.adsabs.harvard.edu/abs/2012ApJ...749L..35U}
}

@ARTICLE{Yoshiike2013ApJ...768..179Y,
       author = {{Yoshiike}, S. and {Fukuda}, T. and {Sano}, H. and {Ohama}, A. and {Moribe}, N. and {Torii}, K. and {Hayakawa}, T. and {Okuda}, T. and {Yamamoto}, H. and {Tajima}, H. and {Mizuno}, N. and {Nishimura}, A. and {Kimura}, K. and {Maezawa}, H. and {Onishi}, T. and {Mizuno}, A. and {Ogawa}, H. and {Giuliani}, A. and {Koo}, B. -C. and {Fukui}, Y.},
        title = "{The Neutral Interstellar Gas toward SNR W44: Candidates for Target Protons in Hadronic {\ensuremath{\gamma}}-Ray Production in a Middle-aged Supernova Remnant}",
      journal = {\apj},
         year = 2013,
        month = may,
       volume = {768},
       number = {2},
          eid = {179},
        pages = {179},
          doi = {10.1088/0004-637X/768/2/179},
       adsurl = {https://ui.adsabs.harvard.edu/abs/2013ApJ...768..179Y}
}

@ARTICLE{Fukui2012ApJ...746...82F,
       author = {{Fukui}, Y. and {Sano}, H. and {Sato}, J. and {Torii}, K. and {Horachi}, H. and {Hayakawa}, T. and {McClure-Griffiths}, N.~M. and {Rowell}, G. and {Inoue}, T. and {Inutsuka}, S. and {Kawamura}, A. and {Yamamoto}, H. and {Okuda}, T. and {Mizuno}, N. and {Onishi}, T. and {Mizuno}, A. and {Ogawa}, H.},
        title = "{A Detailed Study of the Molecular and Atomic Gas toward the {\ensuremath{\gamma}}-Ray Supernova Remnant RX J1713.7-3946: Spatial TeV {\ensuremath{\gamma}}-Ray and Interstellar Medium Gas Correspondence}",
      journal = {\apj},
         year = 2012,
        month = feb,
       volume = {746},
       number = {1},
          eid = {82},
        pages = {82},
          doi = {10.1088/0004-637X/746/1/82},
archivePrefix = {arXiv},
       eprint = {1107.0508},
 primaryClass = {astro-ph.GA},
       adsurl = {https://ui.adsabs.harvard.edu/abs/2012ApJ...746...82F}
}

@ARTICLE{Fukui2015ApJ...798....6F,
       author = {{Fukui}, Y. and {Torii}, K. and {Onishi}, T. and {Yamamoto}, H. and {Okamoto}, R. and {Hayakawa}, T. and {Tachihara}, K. and {Sano}, H.},
        title = "{Optically Thick H I Dominant in the Local Interstellar Medium: An Alternative Interpretation to ``Dark Gas''}",
      journal = {\apj},
         year = 2015,
        month = jan,
       volume = {798},
       number = {1},
          eid = {6},
        pages = {6},
          doi = {10.1088/0004-637X/798/1/6},
archivePrefix = {arXiv},
       eprint = {1403.0999},
 primaryClass = {astro-ph.GA},
       adsurl = {https://ui.adsabs.harvard.edu/abs/2015ApJ...798....6F}
}

@ARTICLE{Fukui2017ApJ...850...71F,
       author = {{Fukui}, Y. and {Sano}, H. and {Sato}, J. and {Okamoto}, R. and {Fukuda}, T. and {Yoshiike}, S. and {Hayashi}, K. and {Torii}, K. and {Hayakawa}, T. and {Rowell}, G. and {Filipovi{\'c}}, M.~D. and {Maxted}, N. and {McClure-Griffiths}, N.~M. and {Kawamura}, A. and {Yamamoto}, H. and {Okuda}, T. and {Mizuno}, N. and {Tachihara}, K. and {Onishi}, T. and {Mizuno}, A. and {Ogawa}, H.},
        title = "{A Detailed Study of the Interstellar Protons toward the TeV {\ensuremath{\gamma}}-Ray SNR RX J0852.0-4622 (G266.2-1.2, Vela Jr.): The Third Case of the {\ensuremath{\gamma}}-Ray and ISM Spatial Correspondence}",
      journal = {\apj},
         year = 2017,
        month = nov,
       volume = {850},
       number = {1},
          eid = {71},
        pages = {71},
          doi = {10.3847/1538-4357/aa9219},
archivePrefix = {arXiv},
       eprint = {1708.07911},
 primaryClass = {astro-ph.HE},
       adsurl = {https://ui.adsabs.harvard.edu/abs/2017ApJ...850...71F}
}

@INPROCEEDINGS{Fukui2003IAUS..221P.224F,
       author = {{Fukui}, Yasuo and {Mizuno}, Norikazu and {Onishi}, Toshikazu},
        title = "{Giant Molecular Clouds and Star Formation in the LMC and SMC}",
    booktitle = {IAU Symposium},
         year = 2003,
       series = {IAU Symposium},
       volume = {221},
        month = jan,
        pages = {P224},
       adsurl = {https://ui.adsabs.harvard.edu/abs/2003IAUS..221P.224F}
}

@ARTICLE{Sano2017ApJ...843...61S,
       author = {{Sano}, H. and {Yamane}, Y. and {Voisin}, F. and {Fujii}, K. and {Yoshiike}, S. and {Inaba}, T. and {Tsuge}, K. and {Babazaki}, Y. and {Mitsuishi}, I. and {Yang}, R. and {Aharonian}, F. and {Rowell}, G. and {Filipovi{\'c}}, M.~D. and {Mizuno}, N. and {Tachihara}, K. and {Kawamura}, A. and {Onishi}, T. and {Fukui}, Y.},
        title = "{Discovery of Molecular and Atomic Clouds Associated with the Magellanic Superbubble 30 Doradus C}",
      journal = {\apj},
         year = 2017,
        month = jul,
       volume = {843},
       number = {1},
          eid = {61},
        pages = {61},
          doi = {10.3847/1538-4357/aa73e0},
archivePrefix = {arXiv},
       eprint = {1701.01962},
 primaryClass = {astro-ph.GA},
       adsurl = {https://ui.adsabs.harvard.edu/abs/2017ApJ...843...61S}
}

@ARTICLE{Yang2014A&A...567A..23Y,
       author = {{Yang}, Rui-zhi and {Zhang}, Xiao and {Yuan}, Qiang and {Liu}, Siming},
        title = "{Fermi Large Area Telescope observations of the supernova remnant HESS J1731-347}",
      journal = {\aap},
         year = 2014,
        month = jul,
       volume = {567},
          eid = {A23},
        pages = {A23},
          doi = {10.1051/0004-6361/201322737},
archivePrefix = {arXiv},
       eprint = {1405.4888},
 primaryClass = {astro-ph.HE},
       adsurl = {https://ui.adsabs.harvard.edu/abs/2014A&A...567A..23Y}
}

@ARTICLE{Lee2013ApJ...767...20L,
       author = {{Lee}, Shiu-Hang and {Slane}, Patrick O. and {Ellison}, Donald C. and {Nagataki}, Shigehiro and {Patnaude}, Daniel J.},
        title = "{A CR-hydro-NEI Model of Multi-wavelength Emission from the Vela Jr. Supernova Remnant (SNR RX J0852.0-4622)}",
      journal = {\apj},
         year = 2013,
        month = apr,
       volume = {767},
       number = {1},
          eid = {20},
        pages = {20},
          doi = {10.1088/0004-637X/767/1/20},
archivePrefix = {arXiv},
       eprint = {1302.4645},
 primaryClass = {astro-ph.HE},
       adsurl = {https://ui.adsabs.harvard.edu/abs/2013ApJ...767...20L}
}

@ARTICLE{Ellison2010ApJ...712..287E,
       author = {{Ellison}, Donald C. and {Patnaude}, Daniel J. and {Slane}, Patrick and {Raymond}, John},
        title = "{Efficient Cosmic Ray Acceleration, Hydrodynamics, and Self-Consistent Thermal X-Ray Emission Applied to Supernova Remnant RX J1713.7-3946}",
      journal = {\apj},
         year = 2010,
        month = mar,
       volume = {712},
       number = {1},
        pages = {287-293},
          doi = {10.1088/0004-637X/712/1/287},
archivePrefix = {arXiv},
       eprint = {1001.1932},
 primaryClass = {astro-ph.HE},
       adsurl = {https://ui.adsabs.harvard.edu/abs/2010ApJ...712..287E}
}

@INPROCEEDINGS{Gabici2013ASSP...34..221G,
       author = {{Gabici}, Stefano},
        title = "{Cosmic Rays and Molecular Clouds}",
    booktitle = {Cosmic Rays in Star-Forming Environments},
         year = 2013,
       editor = {{Torres}, Diego F. and {Reimer}, Olaf},
       series = {Astrophysics and Space Science Proceedings},
       volume = {34},
        month = jan,
        pages = {221},
          doi = {10.1007/978-3-642-35410-6_16},
archivePrefix = {arXiv},
       eprint = {1208.4979},
 primaryClass = {astro-ph.HE},
       adsurl = {https://ui.adsabs.harvard.edu/abs/2013ASSP...34..221G}
}

@ARTICLE{Drury1983RPPh...46..973D,
       author = {{Drury}, L. Oc.},
        title = "{REVIEW ARTICLE: An introduction to the theory of diffusive shock acceleration of energetic particles in tenuous plasmas}",
      journal = {Reports on Progress in Physics},
         year = 1983,
        month = aug,
       volume = {46},
       number = {8},
        pages = {973-1027},
          doi = {10.1088/0034-4885/46/8/002},
       adsurl = {https://ui.adsabs.harvard.edu/abs/1983RPPh...46..973D}
}

@ARTICLE{Blandford1978ApJ...221L..29B,
       author = {{Blandford}, R.~D. and {Ostriker}, J.~P.},
        title = "{Particle acceleration by astrophysical shocks.}",
      journal = {\apjl},
         year = 1978,
        month = apr,
       volume = {221},
        pages = {L29-L32},
          doi = {10.1086/182658},
       adsurl = {https://ui.adsabs.harvard.edu/abs/1978ApJ...221L..29B}
}

@ARTICLE{bell1978MNRAS.182..147B,
       author = {{Bell}, A.~R.},
        title = "{The acceleration of cosmic rays in shock fronts - I.}",
      journal = {\mnras},
         year = 1978,
        month = jan,
       volume = {182},
        pages = {147-156},
          doi = {10.1093/mnras/182.2.147},
       adsurl = {https://ui.adsabs.harvard.edu/abs/1978MNRAS.182..147B}
}

@ARTICLE{Fukuda_2014,
doi = {10.1088/0004-637X/788/1/94},
url = {https://dx.doi.org/10.1088/0004-637X/788/1/94},
year = {2014},
month = {may},
publisher = {The American Astronomical Society},
volume = {788},
number = {1},
pages = {94},
author = {Fukuda, T. and Yoshiike, S. and Sano, H. and Torii, K. and Yamamoto, H. and Acero, F. and Fukui, Y.},
title = {INTERSTELLAR PROTONS IN THE TeV γ-RAY SNR HESS J1731-347: POSSIBLE EVIDENCE FOR THE COEXISTENCE OF HADRONIC AND LEPTONIC γ-RAYS},
journal = {The Astrophysical Journal}
}

@ARTICLE{astropy_2013,
       author = {{Astropy Collaboration} and {Robitaille}, Thomas P. and {Tollerud}, Erik J. and {Greenfield}, Perry and {Droettboom}, Michael and {Bray}, Erik and {Aldcroft}, Tom and {Davis}, Matt and {Ginsburg}, Adam and {Price-Whelan}, Adrian M. and {Kerzendorf}, Wolfgang E. and {Conley}, Alexander and {Crighton}, Neil and {Barbary}, Kyle and {Muna}, Demitri and {Ferguson}, Henry and {Grollier}, Fr{\'e}d{\'e}ric and {Parikh}, Madhura M. and {Nair}, Prasanth H. and {Unther}, Hans M. and {Deil}, Christoph and {Woillez}, Julien and {Conseil}, Simon and {Kramer}, Roban and {Turner}, James E.~H. and {Singer}, Leo and {Fox}, Ryan and {Weaver}, Benjamin A. and {Zabalza}, Victor and {Edwards}, Zachary I. and {Azalee Bostroem}, K. and {Burke}, D.~J. and {Casey}, Andrew R. and {Crawford}, Steven M. and {Dencheva}, Nadia and {Ely}, Justin and {Jenness}, Tim and {Labrie}, Kathleen and {Lim}, Pey Lian and {Pierfederici}, Francesco and {Pontzen}, Andrew and {Ptak}, Andy and {Refsdal}, Brian and {Servillat}, Mathieu and {Streicher}, Ole},
        title = "{Astropy: A community Python package for astronomy}",
      journal = {\aap},
         year = 2013,
        month = oct,
       volume = {558},
          eid = {A33},
        pages = {A33},
          doi = {10.1051/0004-6361/201322068},
archivePrefix = {arXiv},
       eprint = {1307.6212},
 primaryClass = {astro-ph.IM},
       adsurl = {https://ui.adsabs.harvard.edu/abs/2013A&A...558A..33A}
}

@ARTICLE{fukui2024ApJ...961..162F,
       author = {{Fukui}, Yasuo and {Aruga}, Maki and {Sano}, Hidetoshi and {Hayakawa}, Takahiro and {Inoue}, Tsuyoshi and {Rowell}, Gavin and {Einecke}, Sabrina and {Tachihara}, Kengo},
        title = "{The Gamma-Ray Origin of RX J0852.0-4622 Quantifying the Hadronic and Leptonic Components: Further Evidence for the Cosmic-Ray Acceleration in Young Shell-type SNRs}",
      journal = {\apj},
         year = 2024,
        month = feb,
       volume = {961},
       number = {2},
          eid = {162},
        pages = {162},
          doi = {10.3847/1538-4357/ad0da3},
archivePrefix = {arXiv},
       eprint = {2311.11355},
 primaryClass = {astro-ph.HE},
       adsurl = {https://ui.adsabs.harvard.edu/abs/2024ApJ...961..162F}
}

@ARTICLE{aruga2022ApJ...938...94A,
       author = {{Aruga}, M. and {Sano}, H. and {Fukui}, Y. and {Reynoso}, E.~M. and {Rowell}, G. and {Tachihara}, K.},
        title = "{Molecular and Atomic Clouds Associated with the Gamma-Ray Supernova Remnant Puppis A}",
      journal = {\apj},
         year = 2022,
        month = oct,
       volume = {938},
       number = {2},
          eid = {94},
        pages = {94},
          doi = {10.3847/1538-4357/ac90c6},
archivePrefix = {arXiv},
       eprint = {2206.00211},
 primaryClass = {astro-ph.HE},
       adsurl = {https://ui.adsabs.harvard.edu/abs/2022ApJ...938...94A}
}

@ARTICLE{sano2022ApJ...933..157S,
       author = {{Sano}, H. and {Yamaguchi}, H. and {Aruga}, M. and {Fukui}, Y. and {Tachihara}, K. and {Filipovi{\'c}}, M.~D. and {Rowell}, G.},
        title = "{An Expanding Shell of Neutral Hydrogen Associated with SN 1006: Hints for the Single-degenerate Origin and Faint Hadronic Gamma-Rays}",
      journal = {\apj},
         year = 2022,
        month = jul,
       volume = {933},
       number = {2},
          eid = {157},
        pages = {157},
          doi = {10.3847/1538-4357/ac7465},
archivePrefix = {arXiv},
       eprint = {2205.13712},
 primaryClass = {astro-ph.HE},
       adsurl = {https://ui.adsabs.harvard.edu/abs/2022ApJ...933..157S}
}

@ARTICLE{sano2021ApJ...919..123S,
       author = {{Sano}, H. and {Yoshiike}, S. and {Yamane}, Y. and {Hayashi}, K. and {Enokiya}, R. and {Tokuda}, K. and {Tachihara}, K. and {Rowell}, G. and {Filipovi{\'c}}, M.~D. and {Fukui}, Y.},
        title = "{ALMA CO Observations of the Mixed-morphology Supernova Remnant W49B: Efficient Production of Recombining Plasma and Hadronic Gamma Rays via Shock-Cloud Interactions}",
      journal = {\apj},
         year = 2021,
        month = oct,
       volume = {919},
       number = {2},
          eid = {123},
        pages = {123},
          doi = {10.3847/1538-4357/ac0dba},
archivePrefix = {arXiv},
       eprint = {2106.12009},
 primaryClass = {astro-ph.HE},
       adsurl = {https://ui.adsabs.harvard.edu/abs/2021ApJ...919..123S}
}

@ARTICLE{stil2006AJ....132.1158S,
       author = {{Stil}, J.~M. and {Taylor}, A.~R. and {Dickey}, J.~M. and {Kavars}, D.~W. and {Martin}, P.~G. and {Rothwell}, T.~A. and {Boothroyd}, A.~I. and {Lockman}, Felix J. and {McClure-Griffiths}, N.~M.},
        title = "{The VLA Galactic Plane Survey}",
      journal = {\aj},
         year = 2006,
        month = sep,
       volume = {132},
       number = {3},
        pages = {1158-1176},
          doi = {10.1086/505940},
archivePrefix = {arXiv},
       eprint = {astro-ph/0605422},
 primaryClass = {astro-ph},
       adsurl = {https://ui.adsabs.harvard.edu/abs/2006AJ....132.1158S}
}
\bibliographystyle{aasjournal}

%% the collaboration and author truncationcommands are used.  It has to


\end{document}